\def\hi{H~{\sc i}}

\documentclass[useAMS,usenatbib,epsfig]{mn2e}

\usepackage{rotating}
\usepackage{float,amssymb}
\usepackage{natbib}
\usepackage{verbatim}
\usepackage{hyperref,graphicx}

\title[GRB X-shooter extinction curves]{VLT/X-shooter GRBs: Individual extinction curves of star-forming regions\thanks{Based on the spectroscopic observations collected at the European Organisation for Astronomical Research in the Southern Hemisphere, 8.2 m Very Large Telescope (VLT) with the X-shooter instrument mounted at UT2 under ESO programmes 060.A-9022(C), 060.A-9427(A), 084.A-0260(B), 085.A-0009(B), 086.A-0073(B), 088.A-0051(B), 089.A-0067(B), 0.90.A-0088(B), 091.C-0934(C), and 092.A-0124(A).}} 
\author[T. Zafar et al.] {T. Zafar$^{1, 2}$\thanks{e-mail:tayyaba.zafar@aao.gov.au}, D. Watson$^3$, P. M{\o}ller$^2$, J. Selsing$^3$, J. P. U. Fynbo$^3$, P. Schady$^4$,
\newauthor
K. Wiersema$^5$, A. J. Levan$^5$, K. E. Heintz$^{6, 3}$, A. de Ugarte Postigo$^{7, 3}$, V. D'Elia$^{8, 9}$,
\newauthor
P. Jakobsson$^6$, J. Bolmer$^{10}$, J. Japelj$^{11}$, S. Covino$^{12}$, A. Gomboc$^{13}$, and Z. Cano$^{6}$ \\
$^1$ Australian Astronomical Observatory, PO Box 915, North Ryde, NSW 1670, Australia. \\
$^2$ European Southern Observatory, Karl-Schwarzschild-Strasse 2, 85748, Garching, Germany. \\
$^3$ Dark Cosmology Centre, Niels Bohr Institute, University of Copenhagen,
Juliane Maries Vej 30, DK-2100 Copenhagen, Denmark. \\
$^4$ Max-Planck Institut f$\ddot{\rm u}$r Extraterrestrische Physik, Giessenbachstrsse 1, 85748 Garching, Germany. \\
$^5$ Department of Physics, University of Warwick, Coventry, CV4 7AL, UK.\\
$^6$ Centre for Astrophysics and Cosmology, Science Institute, University of Iceland, Dunhagi 5, 107 Reykjav\'ik, Iceland.\\
$^{7}$ Instituto de Astrof\'{i}sica de Andaluc\'{i}a (IAA-CSIC), Glorieta de la Astronomía s/n, E-18008 Granada, Spain.\\
$^8$ Space Science Data Center - Agenzia Spaziale Italiana , Via del Politecnico snc., I-00133 Roma, Italy.\\
$^9$ Istituto Nazionale di Astrofisica -- Osservatorio Astronomico di Roma, Via Frascati 33, I-00040 Monteporzio Catone, Italy.\\
$^{10}$ European Southern Observatory, Alonso de C\'{o}rdova 3107, Vitacura, Casilla 19001, Santiago 19, Chile.\\
$^{11}$ Astronomical Institute Anton Pannekoek, University of Amsterdam, Science Park 904, 1098 XH, Amsterdam, the Netherlands.\\
$^{12}$ Osservatorio Astronomico di Brera, via Bianchi 46, 23807, Merate (LC), Italy.\\
$^{13}$ Centre for Astrophysics and Cosmology, University of Nova Gorica, Vipavska 11c, 5270 Ajdov\v s\v cina, Slovenia.
}

\begin{document}

%\date{Accepted 1988 December 15. Received 1988 December 14; in original form 1988 October 11}

\pagerange{\pageref{firstpage}--\pageref{lastpage}} \pubyear{2018}

\maketitle

\label{firstpage}

\begin{abstract}
The extinction profiles in Gamma-Ray Burst (GRB) afterglow spectral energy distributions (SEDs) are usually described by the Small Magellanic Cloud (SMC)-type extinction curve. In different empirical extinction laws, the total-to-selective extinction, $R_V$, is an important quantity because of its relation to dust grain sizes and compositions. We here analyse a sample of 17 GRBs (0.34$<$$z$$<$7.84) where the ultraviolet to near-infrared spectroscopic observations are available through the VLT/X-shooter instrument, giving us an opportunity to fit individual extinction curves of GRBs for the first time. Our sample is compiled on the basis that multi-band photometry is available around the X-shooter observations. The X-shooter data are combined with the \emph{Swift} X-ray data and a single or broken power-law together with a parametric extinction law is used to model the individual SEDs. We find 10 cases with significant dust, where the derived extinction, $A_V$, ranges from 0.1--1.0\,mag. In four of those, the inferred extinction curves are consistent with the SMC curve. The GRB individual extinction curves have a flat $R_V$ distribution with an optimal weighted combined value of $R_V = 2.61\pm0.08$ (for seven broad coverage cases). The `average GRB extinction curve' is similar to, but slightly steeper than the typical SMC, and consistent with the SMC Bar extinction curve at $\sim$$95$\% confidence level. The resultant steeper extinction curves imply populations of small grains, where large dust grains may be destroyed due to GRB activity. Another possibility could be that young age and/or lower metallicities of GRBs environments are responsible for the steeper curves.
\end{abstract}

\section{Introduction}
`Interstellar dust' plays a crucial role in the formation of stars and the evolution and assembly of galaxies. Dust alters the light in the ultraviolet (UV) and optical wavelength ranges through scattering and absorption. Extinction provides an indirect measure of the enrichment process and conditions within an environment. Extinction curve, a standard tool to study extinction as a function of wavelength, strongly depends on the dust grain size distributions and grain compositions \citep{weingartner01,li01,draine03}.

 At cosmological distances long-duration Gamma-Ray Burst (GRB) afterglows offer a unique probe to study dust in star-forming environments \citep[e.g.,][]{stratta07,li08,liang09,zafar10,stratta11,greiner11,zafar11,schady12}. The long-duration GRBs are associated with the deaths of massive stars \citep[e.g.,][]{galama98,hjorth03,starling11} and their spectral emission is synchrotron due to the interaction between the highly relativistic ejecta and surrounding interstellar medium (ISM), as explained by the fireball model \citep[e.g.][]{gehrels09}. When a GRB triggers, the \emph{Neil Gehrels Swift Observatory} \citep{gehrels04} immediately repoints its telescopes and starts observing its X-ray and UV/optical afterglow. The fast response ground-based telescopes also start obtaining photometric and spectroscopic data at multi-wavelengths providing the X-ray to the near-infrared (NIR) spectral energy distributions (SEDs) for GRB afterglows. Adding X-ray data to the GRB SED allows us to improve the constrains on the spectral slope of the afterglow emission and to model the extinction better.

A standard procedure to study dust is to fit the data with empirical extinction laws \citep{cardelli89,fm90,pei92}. The average extinction curves of the Milky Way (MW) and the Large and Small Magellanic Clouds (LMC and SMC) are different from each other due to varying strength of the rest-frame 2175\,\AA\ dust absorption feature in the former two and absence of the feature and UV-steepness in the latter case. For GRBs widely adopted extinction laws are the fixed MW, LMC, and SMC from \citet{pei92} usually proving to be good for the classification of different types of extinction curves. Typically, GRB SEDs prefer fixed SMC featureless extinction law \citep[e.g.,][]{zafar11} from \citet{pei92} with $R_V=2.93$ (total-to-selective extinction). However, in some cases adjustable parametric laws \citep{fm90} are proven to best match the data \citep{ardis09,perley09,schady12,zafar12}. In the parametric extinction laws, the $R_V$ parameter is of particular interest as its small value defines a steep extinction curve and vice versa. The steepness and flatness of the extinction curve are related to the dust grain size distribution and composition \citep{weingartner01} making this quantity pivotal to understand the dust properties. 

With the advancements of various instruments such as the VLT/X-shooter \citep{vernet11}, the spectra of GRB afterglows are now available at multi-wavelengths. X-shooter spectrograph has simultaneous coverage from the UV to the NIR through three spectroscopic arms: UVB (300-550\,nm), VIS (550-1,000\,nm), and NIR (1,000-2,500\,nm). This broader coverage and the availability of the X-ray data provide us with an opportunity to fit the individual extinction curves of GRB afterglows rather than using fixed laws to understand dust properties at higher redshifts.

In \S2 we present our sample selection criteria and provide details about the multi-wavelength data. In \S3 we describe our parametric dust law and SED analysis. Our results are presented in \S4, and a discussion and conclusions are provided in \S5 and \S6. Throughout the paper, errors donate 1$\sigma$ uncertainties unless stated otherwise.

\begin{table}
\begin{minipage}[t]{\columnwidth}
\caption{The X-shooter GRB afterglow sample. The columns indicate: $i)$ GRB name, $ii)$ redshift, $iii)$ Galactic extinction, $iv)$ total Galactic equivalent neutral hydrogen column density, and $v)$ mid-time of the SED.}      
\label{grb:list} 
\centering
\renewcommand{\footnoterule}{}  % to avoid a line before footnotes     
\setlength{\tabcolsep}{9pt}
\begin{tabular}{l c c c c }\hline\hline                       
GRB & $z$ & $E(B-V)_{\rm Gal}$ & $N_{\rm H, Gal}$ & $\Delta t$ \\
	&  & mag &  10$^{20}$ cm$^{-2}$ &  days \\ 
\hline\hline
090313	& 	3.372 & 	0.03 & 2.10  &  1.74  \\
090926A	& 	2.106 & 	0.02 & 2.97  &  0.89  \\
100219A	&	4.667 & 	0.07 & 6.50  &  0.55  \\
100316B	& 	1.180 & 	0.12	& 9.92  &  0.05  \\
100418A	&	0.624 & 	0.06 & 6.08  &  1.47  \\
100814A	& 	1.440 & 	0.02 & 1.85  &  4.10  \\
100901A	& 	1.408 & 	0.09 & 9.49  &  2.76   \\
101219B	& 	0.552 & 	0.02 & 3.32  &  0.48  \\
111008A	& 	4.991 & 	0.01 & 0.99  &  0.41  \\
111209A	& 	0.677 & 	0.02 & 1.54  &  0.90  \\
120119A	& 	1.728 & 	0.10 & 11.3  & 	0.07  \\
120815A	&	2.358 & 	0.10 & 11.9  &   0.08  \\
120923A	&	7.840 & 	0.13 & 15.1  &   0.78  \\
121024A	& 	2.298 & 	0.09 & 7.87  &  0.13  \\
130427A	&	0.339 & 	0.02 & 1.91  & 	0.65  \\	
%130603B  & 	0.356 & 	0.02 & 1.93  &  0.35  \\
130606A  &	5.913 &    0.02 & 2.14  &  0.33  \\
131117A   &	4.042 &	0.02 & 1.50  &  0.02 \\
\hline
\end{tabular}
\end{minipage}
\end{table}

%----------------------------------------------------------------
\section{Data sample}
\subsection{Sample selection}
Under the X-shooter GRB target of opportunity (ToO) programs, spectra for a large sample of GRB afterglows have been acquired from March 2009 to March 2014. The spectra are reduced and flux calibrated using the standard X-shooter pipeline (version 2.0; \citealt{modigliani10}). The X-shooter GRB afterglow spectra are taken usually with slit widths of 1.0$''$, 0.9$''$ and 0.9$''$ for UVB, VIS, and NIR spectra, respectively. A detailed description on the reduction and flux calibration of the spectra will be presented in Selsing et al. (in prep) describing the data reduction including background subtraction, extraction and flux calibration. Our parametric dust extinction analysis is, in particular, sensitive to flux calibration. For the GRB afterglow data, two factors could lead to sub-optimal flux calibration, i.e.: $i)$ The flux standard star observations for X-shooter are taken with a broader 5.0$''$ wide slit and hence has different slit loss than the science spectra, and $ii)$ Atmospheric Dispersion Correctors (ADC) for the UVB and VIS arms were disabled from August 2012 and until recently - after our latest spectra were secured. Therefore, we cannot solely rely on the instrument's response function and we hence require photometric data around each X-shooter observations. We looked into the literature for multi-band photometric data available around each X-shooter observation. We further selected the cases where data is either not affected or corrected for the contamination by the supernova or GRB host galaxy emission. This criterion leaves us with a sample of 17 long-duration GRB afterglows with redshifts ranging from 0.34 $<$ $z$ $<$ 7.84 to conduct the NIR to X-ray SED analysis and derive extinction curves from the star-forming regions.  

Previously, \cite{japelj15} fitted the X-ray to the X-shooter SEDs of nine GRB afterglows using fixed Local Group \citet{pei92} extinction laws and binned X-shooter data. Eight members of our sample overlap with their sample (except short GRB\,130603B), however, we here attempt to fit a free parametric dust model to derive individual extinction curves for the {\it unbinned} X-shooter data for the first time. Our method keeps the original spectral binning of X-shooter and no additional binning during the SED construction process is applied. The overlapping cases are discussed individually in Sect. \ref{grb:det}.

\begin{figure*}
  \centering
{\includegraphics[width=\columnwidth,clip=]{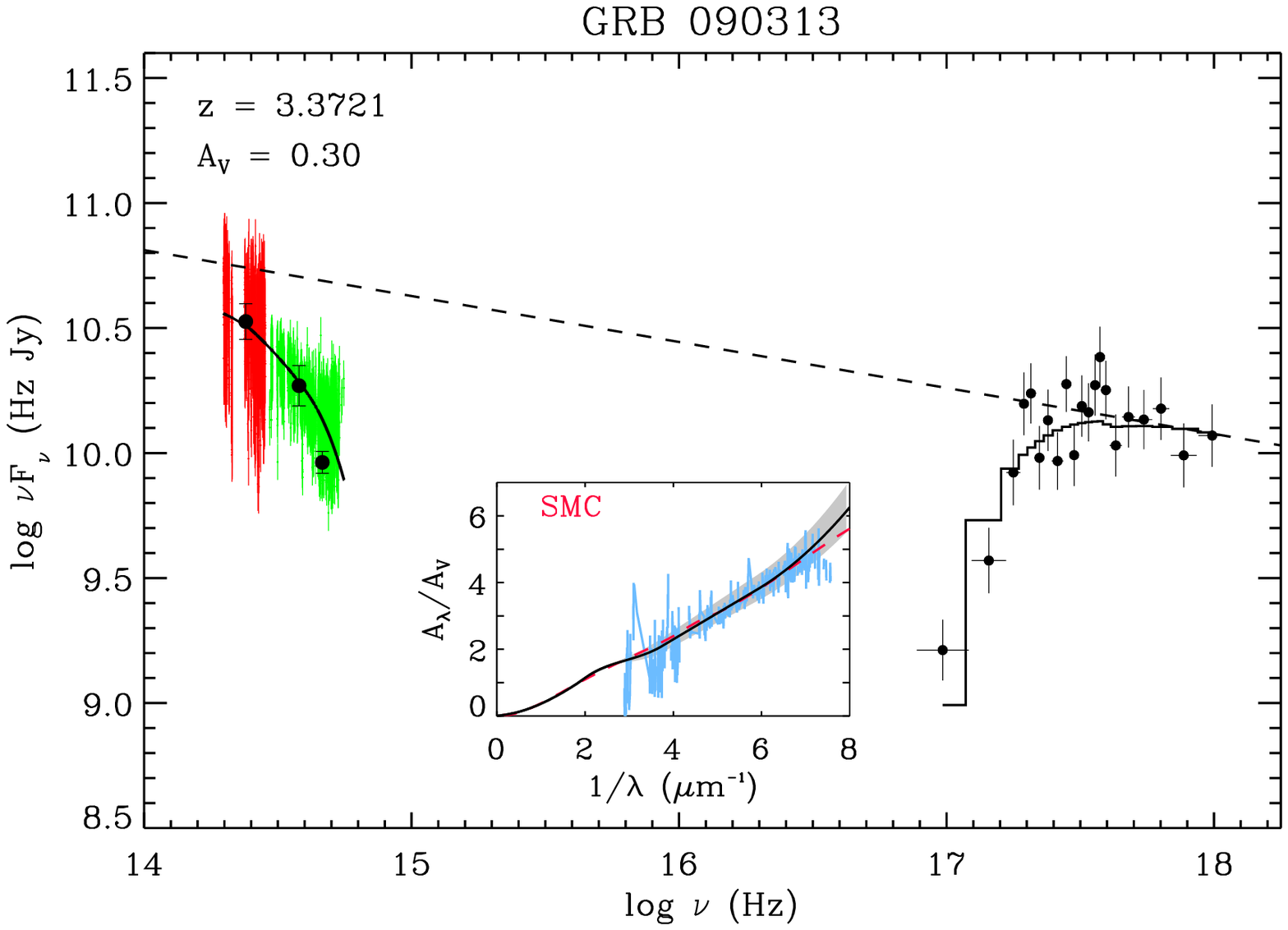}}
 {\includegraphics[width=\columnwidth,clip=]{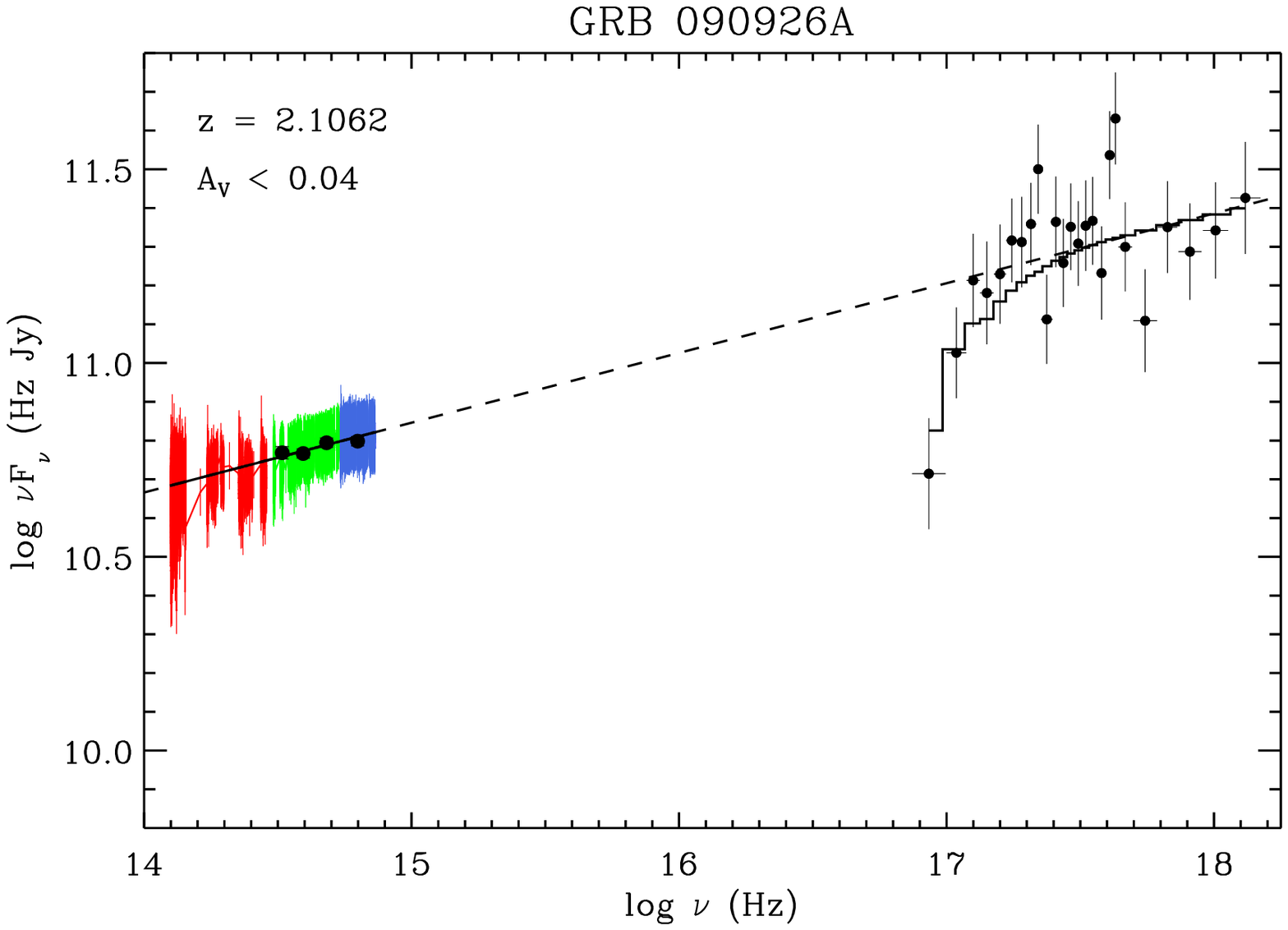}}
{\includegraphics[width=\columnwidth,clip=]{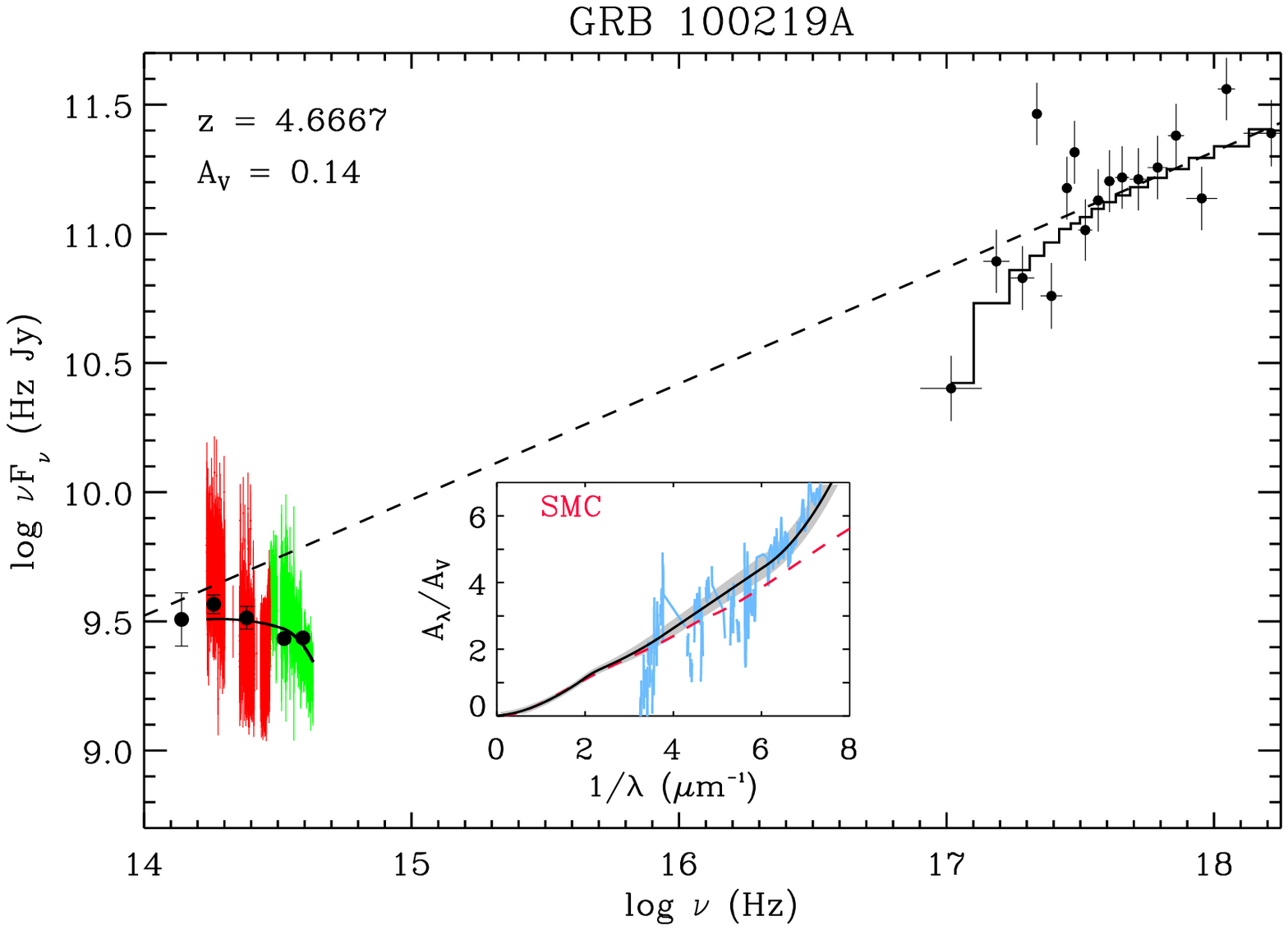}}
{\includegraphics[width=\columnwidth,clip=]{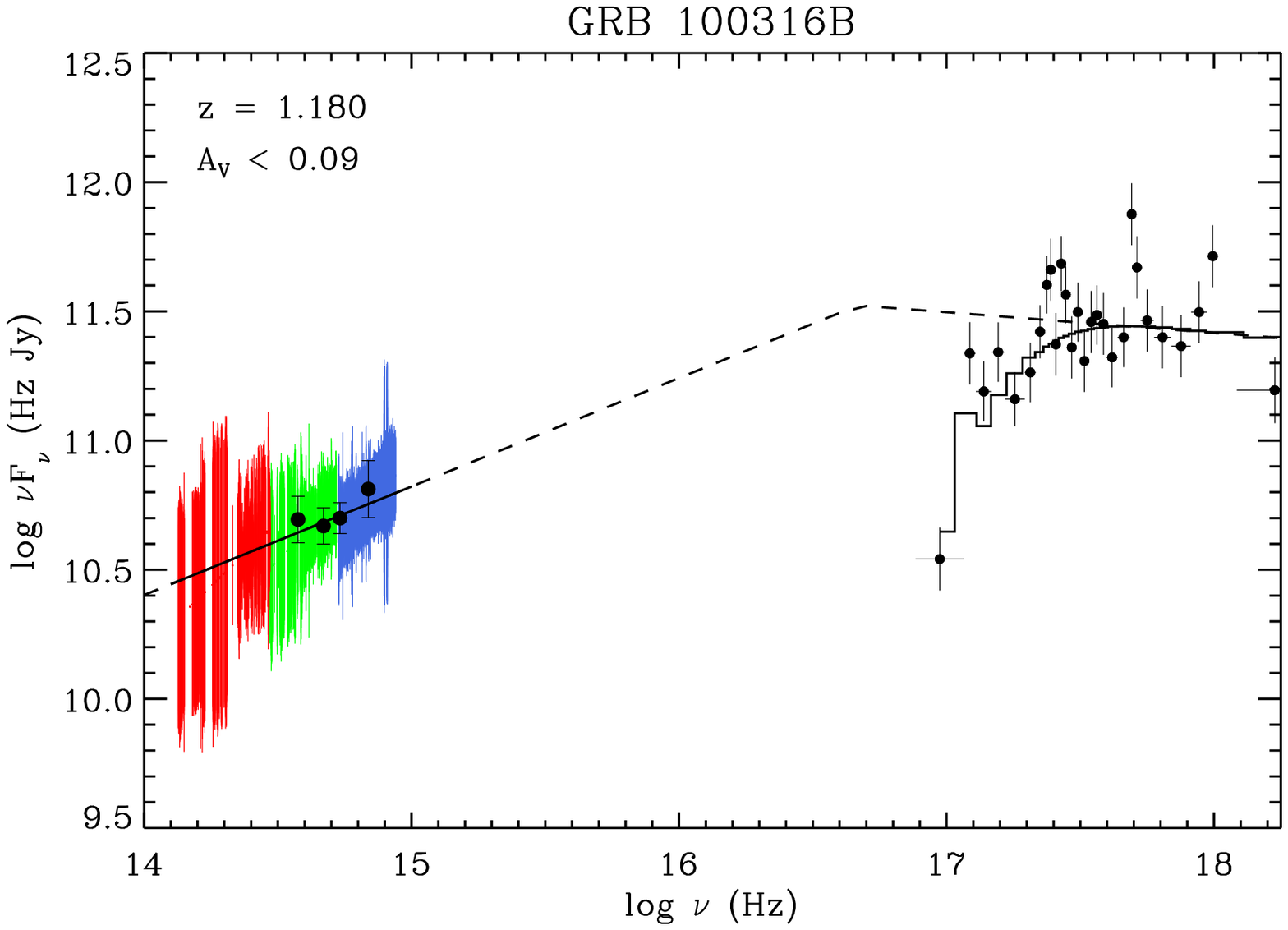}}
{\includegraphics[width=\columnwidth,clip=]{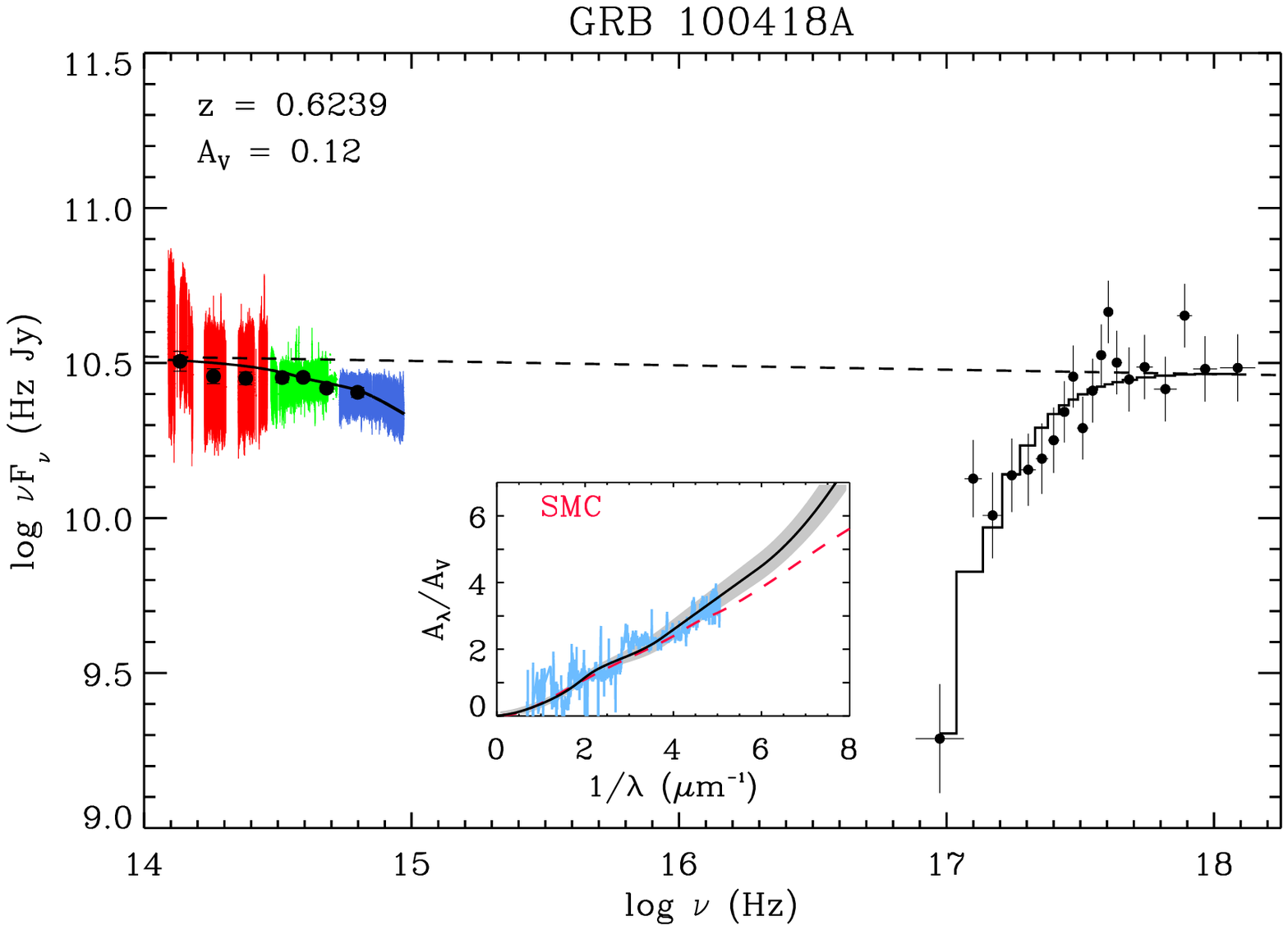}}
{\includegraphics[width=\columnwidth,clip=]{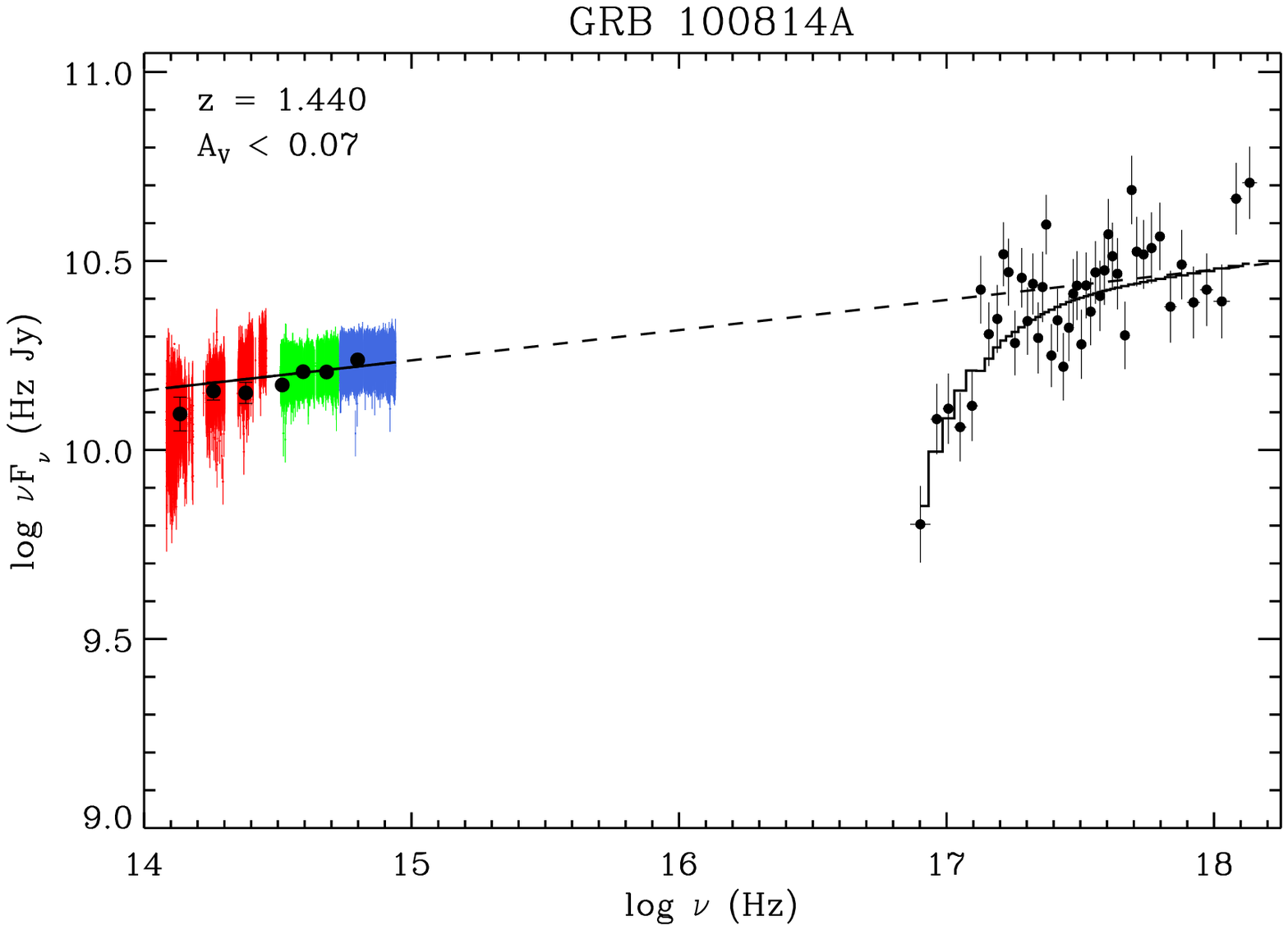}}
     \caption{The observed-frame VLT/X-shooter GRB afterglow SEDs and their best fit models. In each panel, the \emph{Swift} X-ray data on the right side is indicated by black points. The blue, green, and red colours correspond to the UVB, VIS, and NIR spectra from the VLT/X-shooter, respectively. Black overlaid points on the X-shooter spectra are the photometric data from different sources (see \S\ref{grb:det}). The errors on the spectroscopic and photometric data are also plotted. The X-shooter spectra are binned for visual purposes. The best fit extinguished (solid lines) and extinction corrected spectral models (dashed lines) are shown in black. {\it Inset:} For extinguished cases, the best-fit absolute extinction curves of the GRBs are shown with black lines together with their 1$\sigma$ uncertainty with grey shaded area. The cyan curves represent the X-shooter spectra. The red dashed line corresponds to the canonical SMC extinction curve from \citet{pei92}.}
         \label{grb:fits}
  \end{figure*}

\addtocounter{figure}{-1}
\begin{figure*}
  \centering
{\includegraphics[width=\columnwidth,clip=]{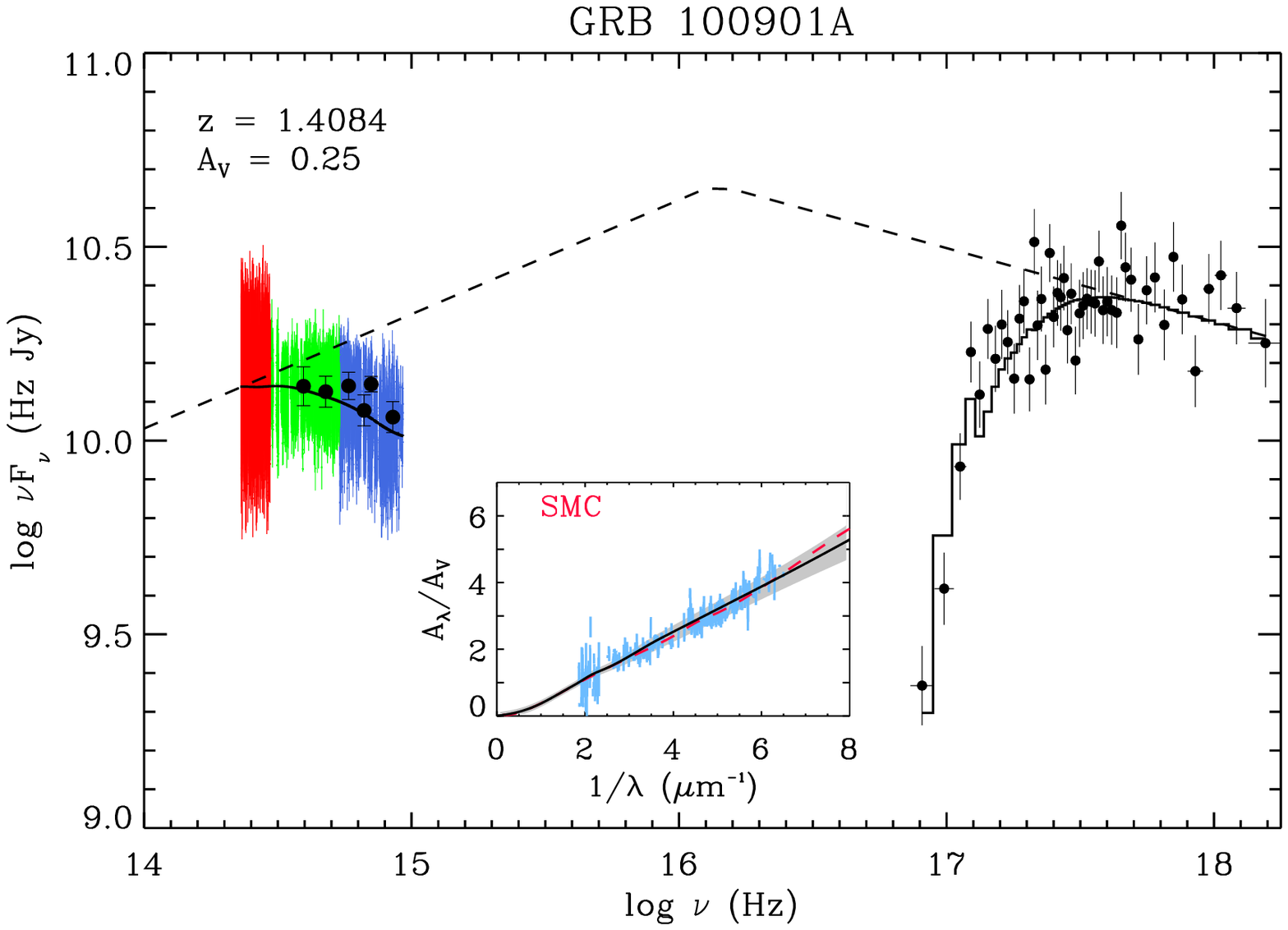}}
{\includegraphics[width=\columnwidth,clip=]{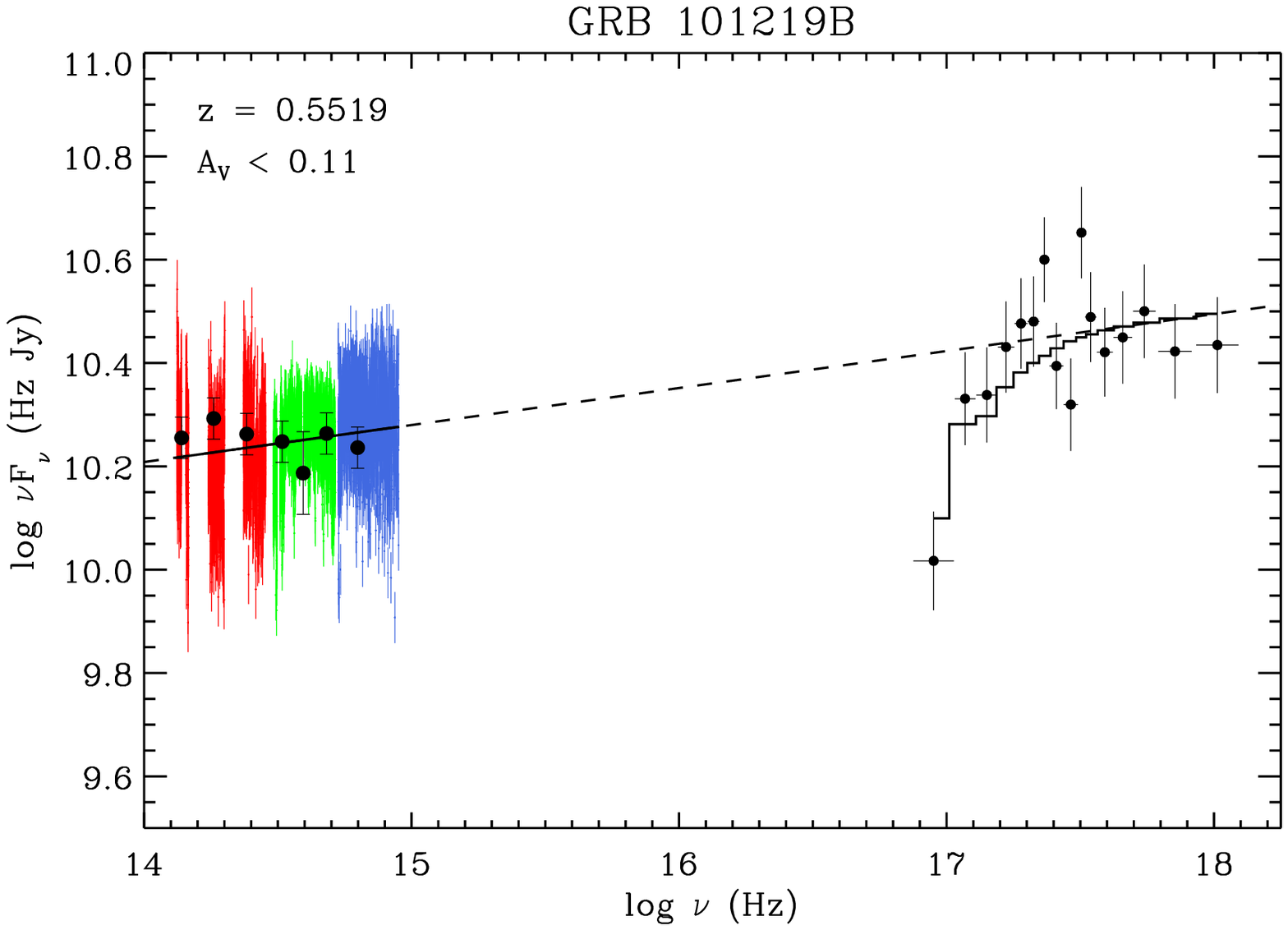}}
{\includegraphics[width=\columnwidth,clip=]{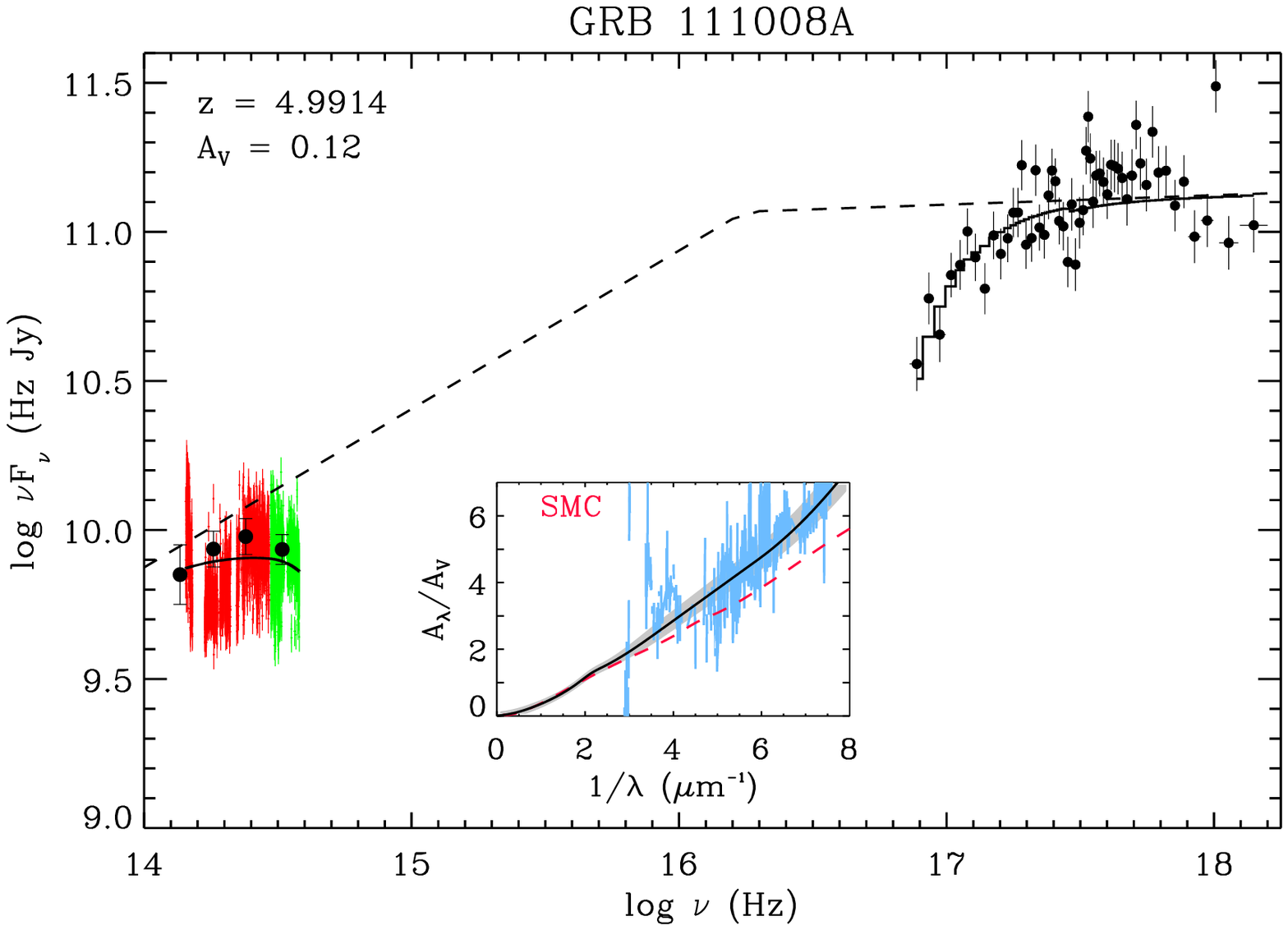}}
{\includegraphics[width=\columnwidth,clip=]{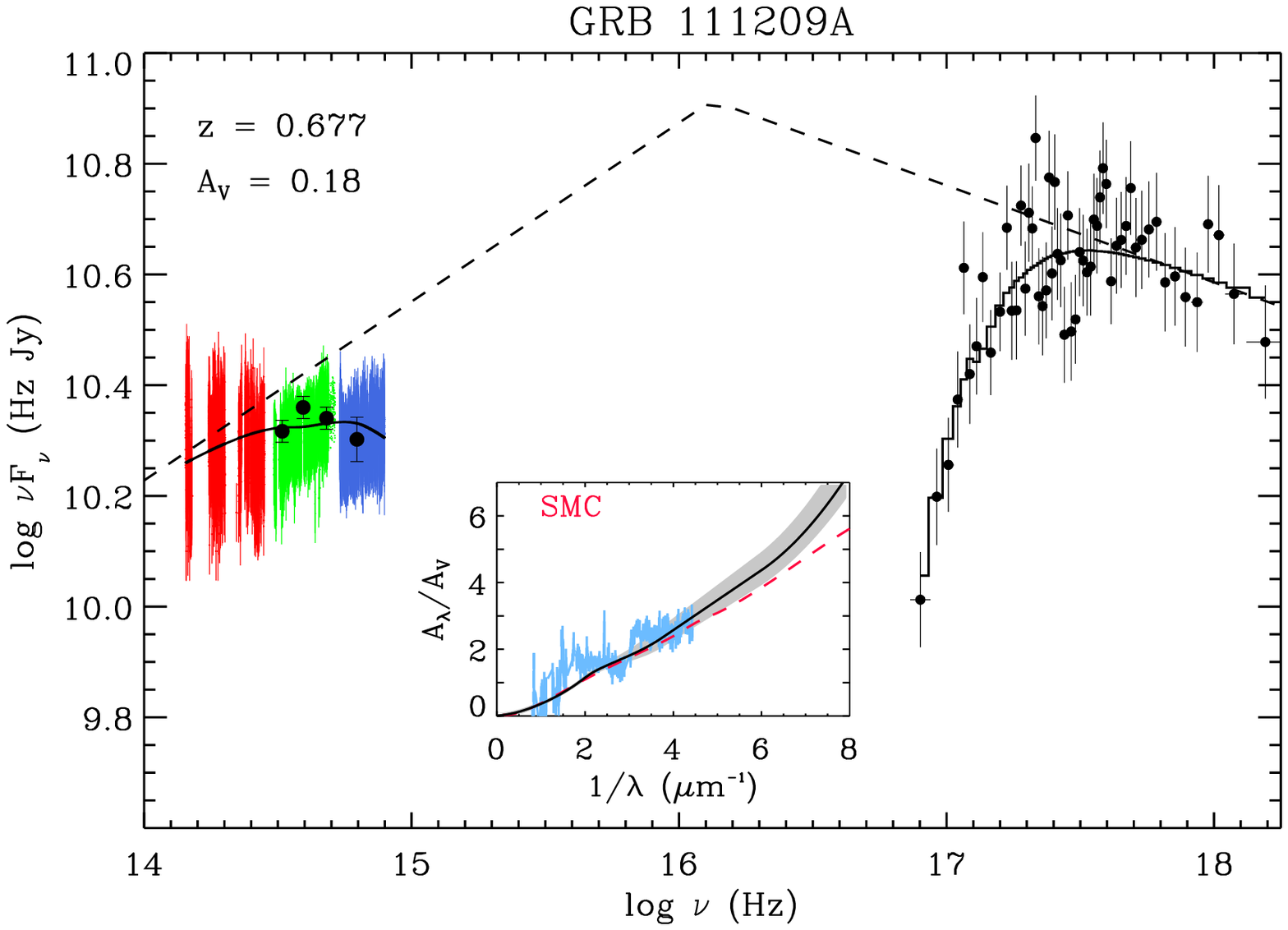}}
{\includegraphics[width=\columnwidth,clip=]{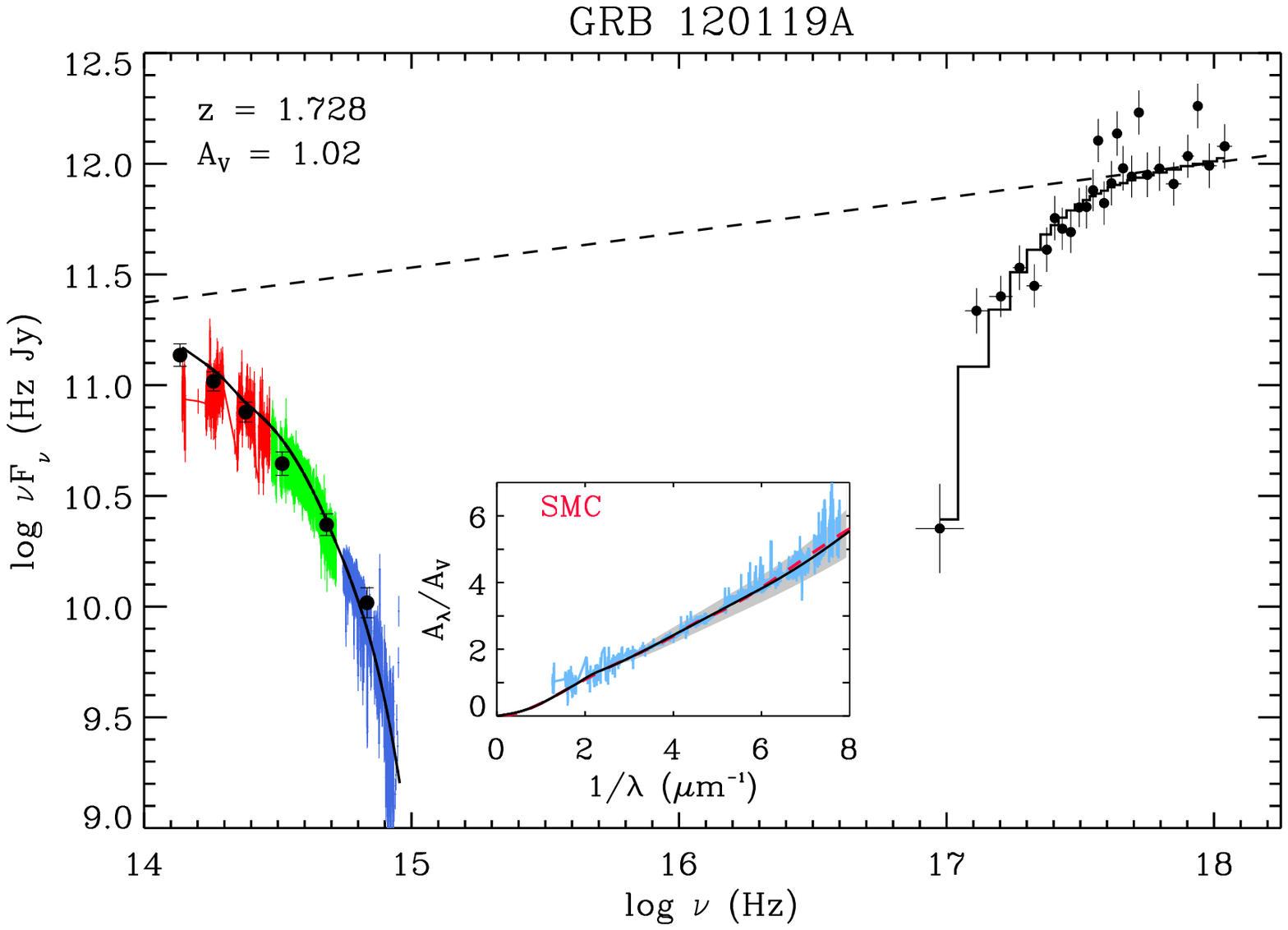}}
{\includegraphics[width=\columnwidth,clip=]{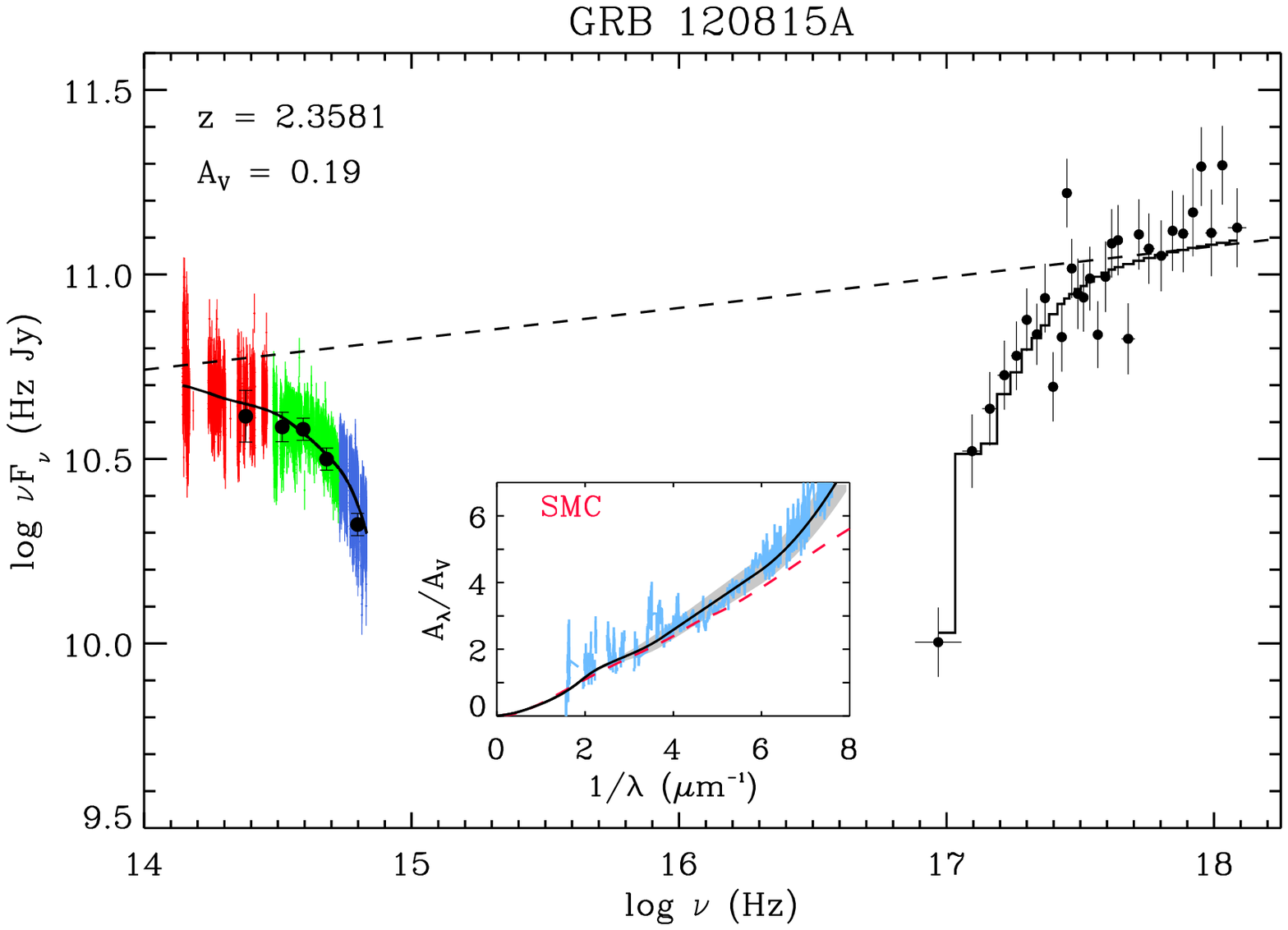}}
{\includegraphics[width=\columnwidth,clip=]{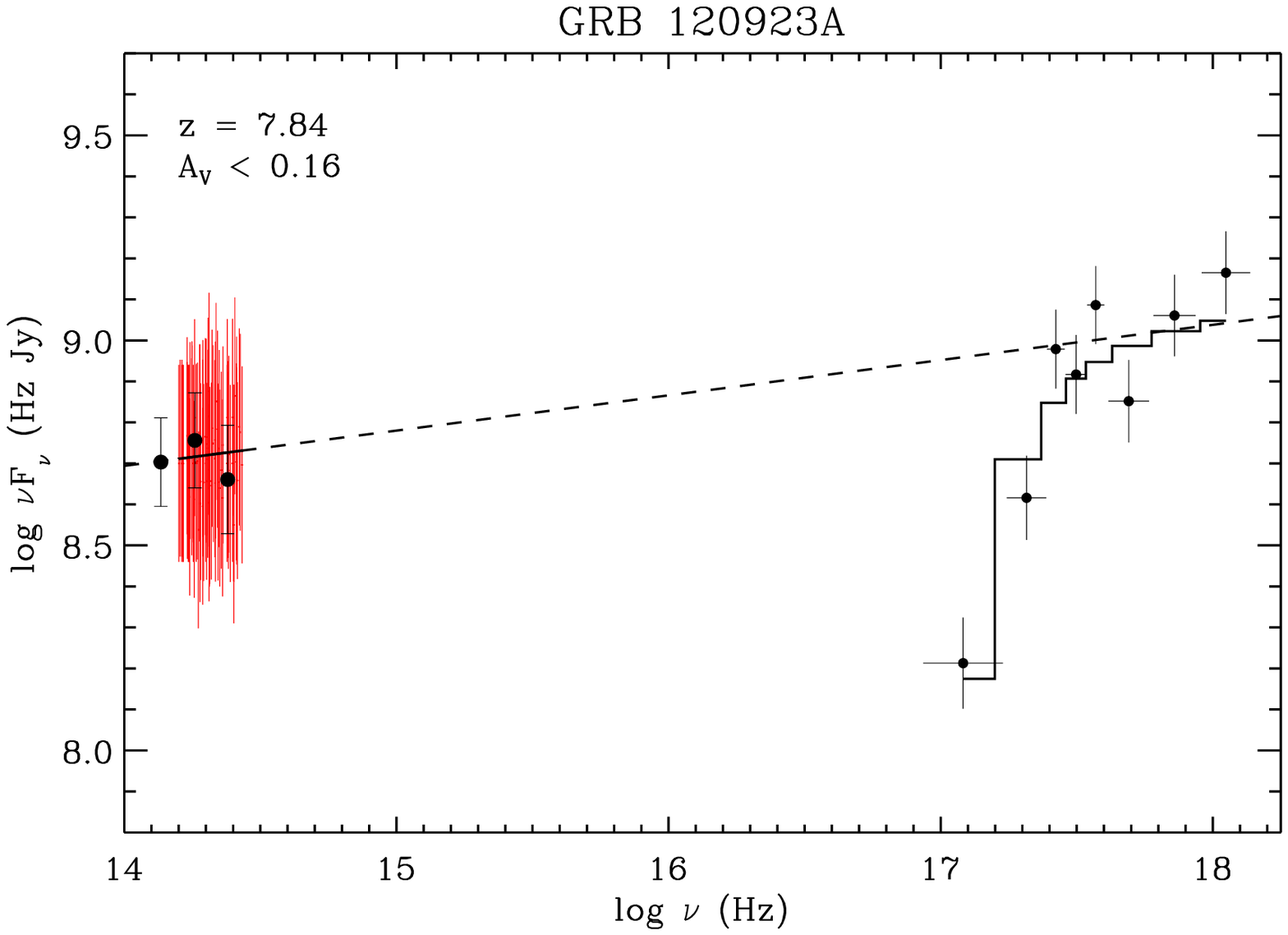}}
{\includegraphics[width=\columnwidth,clip=]{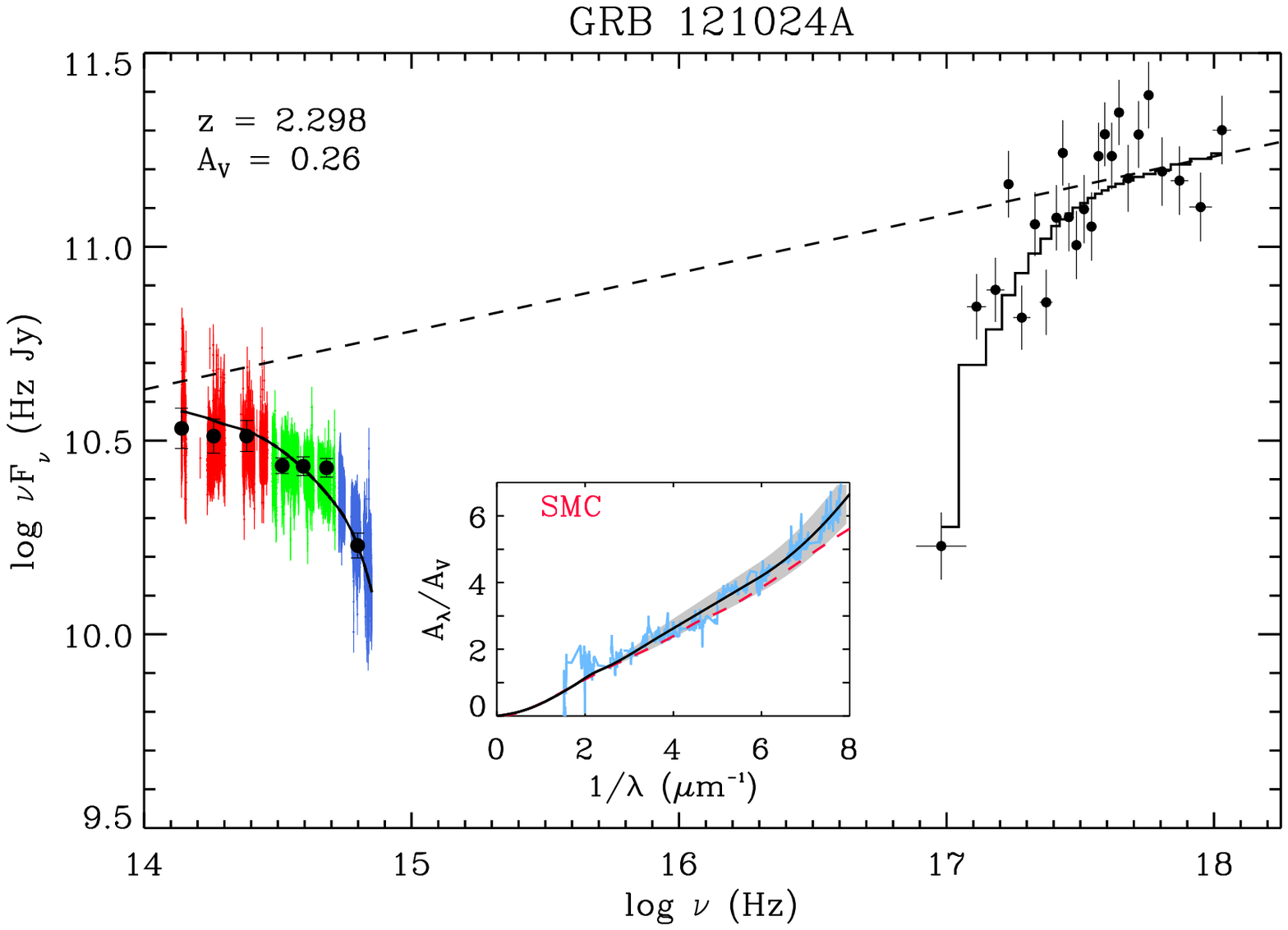}}
%{\includegraphics[width=\columnwidth,clip=]{/Users/tzafar/work/GRB-X-shooter/SED_fit/130603B/SED130603Berc.ps}}
     \caption{Continued.}
  \end{figure*}

\addtocounter{figure}{-1}
\begin{figure*}
  \centering
{\includegraphics[width=\columnwidth,clip=]{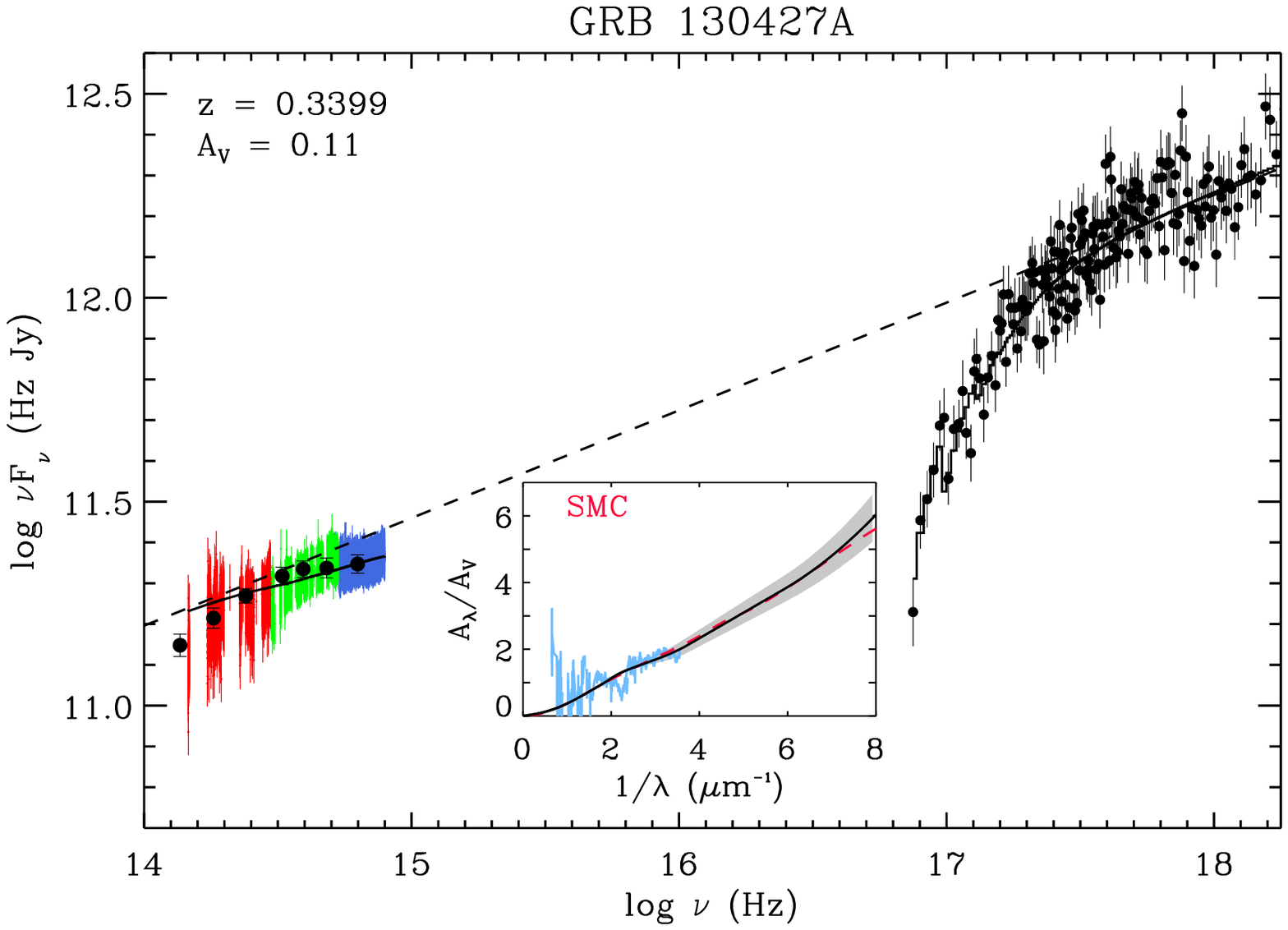}}
{\includegraphics[width=\columnwidth,clip=]{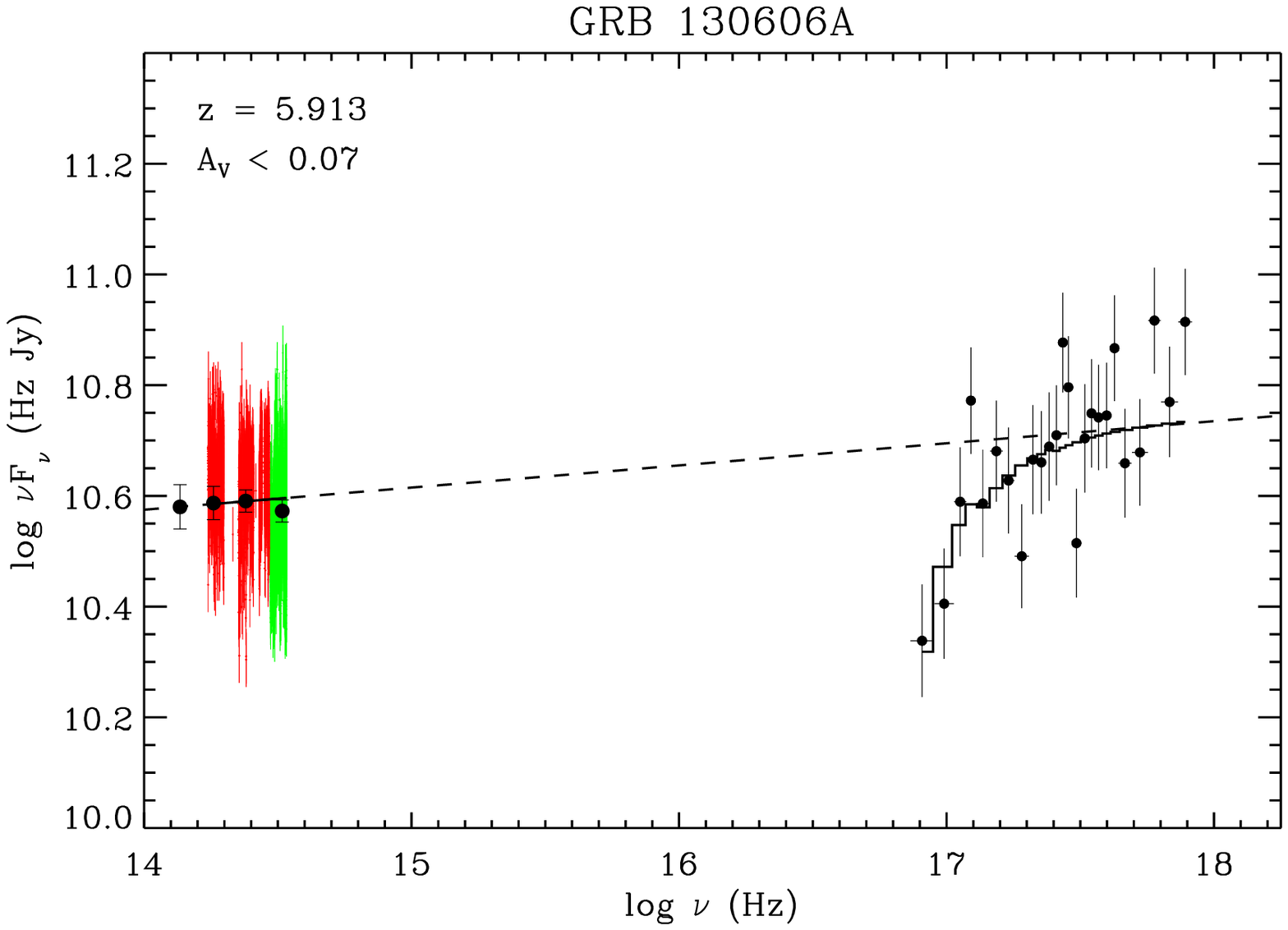}}
  \end{figure*}

\begin{figure}
  \centering
{\includegraphics[width=\columnwidth,clip=]{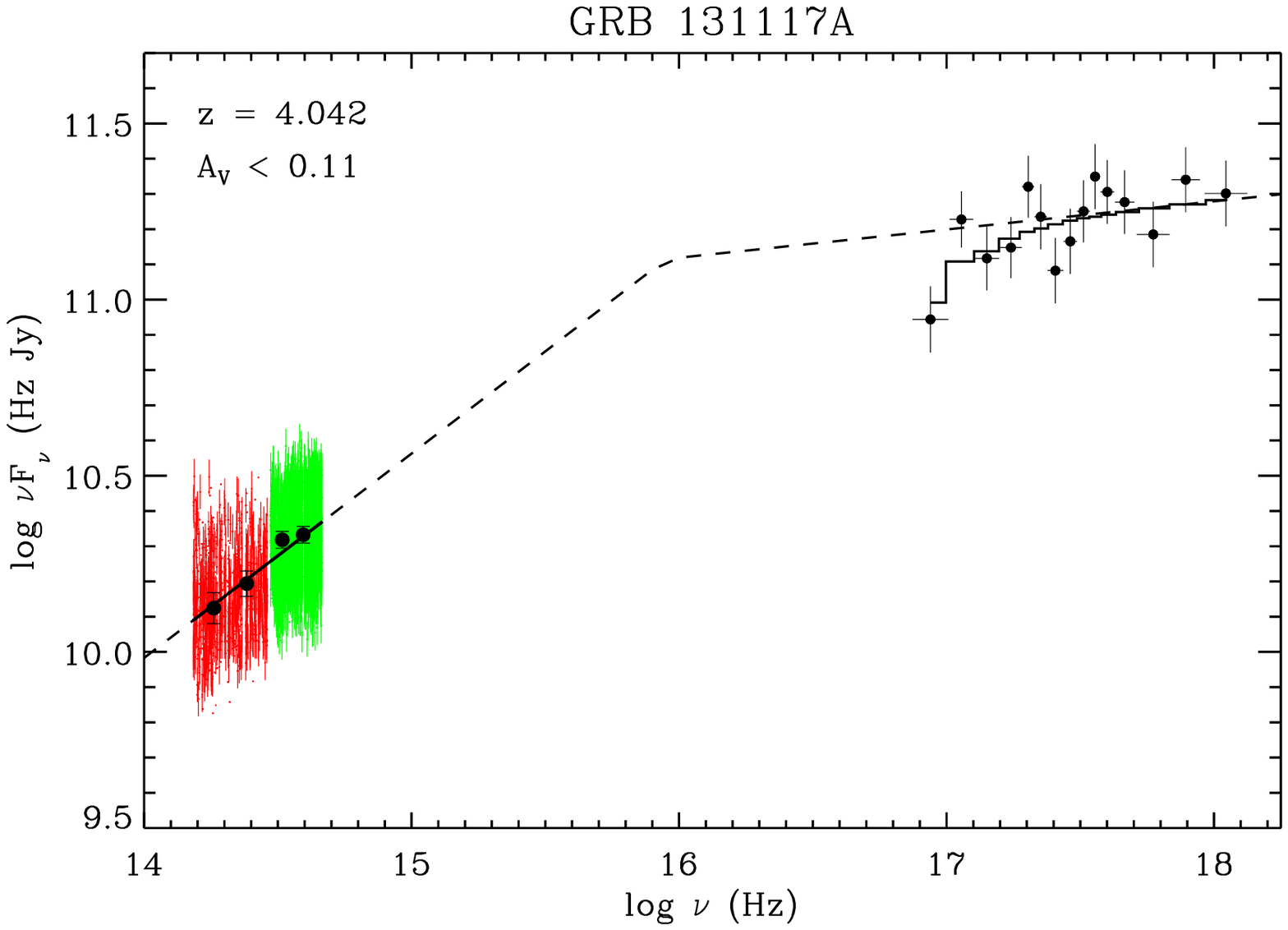}}
     \caption{Continued.}
  \end{figure}

\subsection{X-shooter data}
The X-shooter spectra together with photometric data were corrected for the foreground Galactic extinction (see Table \ref{grb:list}) using the Galactic maps of \citet{schlafly11}. The Galactic extinction is usually small with $E(B-V)\lesssim0.15$, therefore, the uncertainty on this value should have a negligible effect on our final results. The X-shooter spectra were then normalised to the level of the photometric observations to generate SEDs at the photometry mid time, $\Delta t$. The UVB and VIS arm data are comparable with the photometry within $\sim10$\% and the NIR data usually differ by $15-20$\% from the photometry. The HEAsoft software (version 6.19) tool \texttt{flx2xsp} was used to generate XSPEC (version12.9; \citealt{arnaud96}) readable spectral (PHA) and response matrices (RSP) files for the X-shooter data. A background file was then created to mask out non-required data channels. In detail: $i)$ to prevent contamination caused by the damped Ly$\alpha$ absorber and Ly$\alpha$ forest, entire blueward and some redward data (to avoid \hi\ damping wing) around $\lambda_{\rm rest}~<~1216$\,\AA\ is excluded , $ii)$ regions of emission and absorption lines arising from different metal species and atmospheric telluric lines are removed, and $iii)$ bad spikes originating from the sky subtraction residuals (usually in the NIR arm) are masked out to obtain clean continuum data. The PHA, RSP, and background files were then grouped using the \texttt{grppha} tool without any binning applied to individual data channels. 

\subsection{X-ray data}
The \emph{Swift} X-ray Telescope (XRT, \citealt{burrows05}) obtained observations of each GRB afterglow. The X-ray lightcurves were obtained from the \emph{Swift} online repository \citep{evans09} and a decay model \citep{beuermann99} is fitted. The X-ray spectrum in the 0.3--10.0\,keV energy range for each GRB afterglow around the X-shooter spectra mid-time $\Delta t$ was reduced using the HEAsoft software. We verified that the X-ray data show no evidence of spectral evolution so that the lightcurve hardness ratio around $\Delta t$ is not deviating from the mean. We used photon counting (PC) mode data and extracted spectral files using the \texttt{XSELECT} (version 2.4) tool. Response matrices were used from the \emph{Swift} XRT calibration files. The X-ray spectral files were grouped to 20 counts per energy channel using the \texttt{grppha} tool. Using the X-ray lightcurves, the flux of the X-ray data were normalised to the SED mid-time, $\Delta t$, by taking the ratio of the flux level at the SED time and the photon weighted mean time of the X-ray spectrum. For this purpose the \texttt{fparkey} command keyword \texttt{EXPOSURE} is used to correct the 0.3--10.0\,keV flux level.

% \texttt{fparkey} command is used to change the \texttt{EXPOSURE} keyword value of the X-ray data, obtained from the modelled lightcurve flux and flux at the time of the X-shooter data.
%----------------------------------------------------------------
\section{SED analysis}
The rest-frame X-ray through {\it unbinned} optical/NIR instantaneous SEDs of the GRB afterglows were fitted within the spectral fitting package \texttt{XSPEC} using a single or broken power-law together with a parametric extinction law to model for dust. 

The continuum emission from the GRB afterglow is dominated by synchrotron radiation and a single power-law case described as $F_\nu = F_0\nu^{-\beta_1}$. Here $F_0$ is the normalisation flux, $\nu$ is the frequency and $\beta_1$ is the intrinsic spectral slope. In case of a broken power-law model a cooling break, $\nu_{\rm break}$, is introduced and the law is described by two slopes, $\beta_1$ (optical slope) and $\beta_2$ (X-ray slope), as:
\begin{equation}
F_\nu = \left\{ 
\begin{array}{l l}
 F_0 \nu^{-\beta_{1}} & \quad \mbox{ $\nu \leq \nu_{\rm{break}}$ }\\
 F_0 \nu^{ \beta_{2}-\beta_{1}}_{\rm{break}} \nu^{-\beta_2} & \quad \mbox{ $\nu \geq \nu_{\rm{break}}$ } \end{array} \right. 
 \end{equation}
In the latter case, the cooling break was modelled such that the change in slope, $\Delta\beta$, was fixed at 0.5 \citep{sari98}. Such a change in the spectral indices is supported by the analysis of \citet{zafar11} for a spectroscopic sample of GRBs \citep[see also][for a photometric sample analysis]{greiner11}. 

The \texttt{XSPEC} models $phabs$ and $zphabs$ were used to correct the foreground Galactic and host galaxy photoelectric absorptions in the X-ray data. The total Galactic equivalent neutral hydrogen column density, $N_{\rm H, Gal}$, was fixed to the values calculated from \citet{willingale13}. \citet{willingale13} investigated the biases and completeness of the atomic hydrogen column density reported by \citet{kalberla05} and included the contribution from molecular hydrogen \citep{wilms00} and Galactic dust \citep{schlegel98} to that. This results in higher $N_{\rm H, Gal}$ values and carry no systematic errors to our final results. Another prescription to estimate the total Galactic column density based on dust maps was proposed by \citet{watson11}. We use the \citet{willingale13} values because they provide higher Galactic columns. The host galaxy equivalent neutral hydrogen column density from the soft X-ray absorption, $N_{\rm H,X}$, is left as a free parameter. The \texttt{XSPEC} default solar abundances of \citet{anders89} are used following the discussions of \citet{watson11} and \citet{watson13}.

%The \texttt{XSPEC} default solar abundances of \citet{anders89} are used following the discussions of \citet{watson11} and \citet{watson13} due to being a better approximation for a typical Galactic sightline.

\subsection{Dust model}
The optical afterglow is extinguished due to dust absorption and scattering along the line of sight and observed spectra are changed to: $F_{\nu}^{\rm{obs}} = F_\nu10^{-0.4A_\lambda}$, where $A_\lambda$ is the wavelength dependent extinction curve. We use the \citet{fm90} law providing freedom to generate extinction curves with eight free parameters. It contains the galaxy dust extinction, $A_V$, the total-to-selective dust extinction, $R_V$ , and six coefficients defining: $i)$ the UV linear component specified by $c_1$ (intercept) and $c_2$ (slope), $ii)$ the height ($c_3$), width ($\gamma$), and central wavelength ($x_0$) of the 2175\,\AA\ bump, which is modelled with a Lorentzian-like Drude component \citep{bohren83}, $iii)$ and the far-UV curvature term defined by $F(\lambda^{-1})$ and far-UV parameter $c_4$. The extinction properties in the IR and optical ranges are determined using spline interpolation points. The wavelength-dependent extinction is then given as:
\begin{equation}
A_\lambda = \frac{A_V}{R_V}\times\left(c_1+c_2\lambda^{-1}+c_3D(x,x_0,\gamma) +c_4F(\lambda^{-1}) + 1\right)
\end{equation}
Where $F(\lambda^{-1})=$0 for $\lambda^{-1}<5.9$\,$\mu$m$^{-1}$ and $F(\lambda^{-1})=$ $0.5392(\lambda^{-1}-5.9)^2+0.05644(\lambda^{-1}-5.9)^3$ for $\lambda^{-1}\ge5.9$\,$\mu$m$^{-1}$. Hereafter we will refer to this extinction model as FM. We first fit the data with all parameters (including Drude component), but find that not in a single case the $c_3$ parameter is significantly different from zero. This suggests that the 2175\,\AA\ extinction bump is not present for our GRB afterglow sample. We, therefore, fixed the bump parameters $c_3$, $\gamma$, and $x_0$ to zero, 1.0 $\mu m^{-1}$, and 4.6 $\mu m^{-1}$, respectively. This is done to avoid degeneracies of bump parameters ($c_3$, $\gamma$, and $x_0$) when significant bump is not present following the discussions of \citet{zafar15}. \citet{zafar15} tested the SMC-Bar value of $c_3=0.389$ from \citet{gordon03} and $c_3=0$, finding a decrease in $\chi^2$ values with $c_3=0$.

In simultaneous SED analysis within \texttt{XSPEC}, $c_1$, $c_2$, $c_4$, $A_V$, $R_V$, host metal absorption ($N_{\rm H,X}$) and spectral indices of the continuum ($\beta_1$ and $\beta_2$), were fitted as free parameters for each GRB afterglow. The best fit results for each GRB and the resulting $\chi^2$ are provided in the Table \ref{best-fit}. We considered a broken power-law model to provide a better fit to the afterglow SED when the F-test probability is smaller than 5\% . This slightly higher F-test probability threshold is chosen due to relatively higher uncertainties on the X-shooter spectra to avoid the wrong classification of a SED having a single power-law. An incorrect classification could affect the best-fit $A_V$ and hence $R_V$ values. However, for broken power-law cases we always find probability smaller than 1\% .

%----------------------------------------------------------------
\section{Results}\label{grb:det}
We used the X-shooter GRB SEDs to generate individual extinction curves. Note that the photometric data are only used to normalise the X-shooter data to avoid flux calibration discrepancies. The SED fits are only performed on the {\it unbinned} X-shooter and binned XRT data. The best-fit SEDs of GRB afterglows in our sample and extinction curves with their 1$\sigma$ uncertainties for dusty cases (in insets) are shown in Fig. \ref{grb:fits}. As a comparison, we plot the canonical \citet{pei92} SMC extinction curve in the insets. Spectroscopic and photometric data collection, SED generation, and comparison with previous studies are outlined in this section on a case by case basis. 

\subsection{GRB\,090313}
The X-shooter spectra of GRB\,090313 ($z=3.372$) were taken at 1.88 days after the burst trigger during the instrument commissioning. The photometric data at 1.74 days after the burst in the $r^{\prime}I$ and $J$ bands are taken from the lightcurves provided by \citet{melandri10}. The X-shooter spectra were normalised to the optical/NIR photometric data at 1.74 days. The SED fits well with a single power-law and a featureless extinction curve ($R_V=2.67^{+0.13}_{-0.17}$) with $A_V=0.3\pm0.06$. Previously, \citet{kann10} fitted the optical/NIR SED suggesting an SMC-type curve with $A_V=0.34\pm0.15$, consistent to our extinction value.

\subsection{GRB\,090926A}
The X-shooter spectra of the GRB\,090926A ($z=2.016$) afterglow were taken at 0.917 days after the burst trigger. The Gamma-Ray burst Optical and Near-Infrared Detector (GROND) photometry in the $g^{\prime}r^{\prime}i^{\prime}$ and $z^{\prime}$ bands are available from \citet{rau10} at 0.891 days after the burst. Due to technical issues, GROND $JH$ and $K$ band photometry are not available near the epoch of the X-shooter observations. We scaled the X-shooter spectra to the optical photometric data at 0.891 days. The X-shooter to X-ray SED prefers a single power-law and no dust extinction with $A_V<0.04$. Previously \citet{rau10} and \citet{delia10} also reported no dust extinction for this burst with $A_V<0.1$ and $A_V<0.03$, respectively.

\subsection{GRB\,100219A}
The X-shooter UVB to NIR spectra of GRB\,100219A ($z=4.667$) were obtained at 0.55 days after the burst trigger. Photometric data for SED normalisation are obtained from \citet{thoene13} in the $i^{\prime}z^{\prime}JH$ and $K$ bands from GROND. The SED fits well with a single power-law and featureless extinction curve ($R_V=2.65\pm0.09$) with $A_V=0.14\pm0.03$. Previously \citet{thoene13} found best fit with an SMC-type extinction curve and $A_V=0.13\pm0.05$ using the X-shooter spectrum at similar epoch, suggesting consistent results. In contrast, \citet{japelj15} claimed the X-shooter to X-ray SED fits well with an LMC extinction curve with $A_V=0.23\pm0.02$. We do not find any evidence of 2175\,\AA\ the bump (with $c_3<0.26$) in the spectrum and a slightly higher extinction. Recently, \citet{bolmer17} found an SMC-type curve with $A_V=0.15^{+0.04}_{-0.05}$ could explain the GROND-XRT SED, consistent with our results.

\subsection{GRB\,100316B}
At $\sim$0.045 days after the burst, the X-shooter spectra of the GRB\,100316B ($z=1.180$) afterglow were taken. The photometric data for this afterglow are taken from \citet{haislip10} with the Panchromatic Robotic Optical Monitoring and Polarimetry Telescopes (PROMPT) at Cerro Tololo Inter-American Observatory (CTIO) in the $BVR$, and $I$ bands. The X-shooter data were scaled to the photometry at $\sim$0.05 days after the burst. The X-shooter to X-ray SED fits well with a broken power-law (with the break significance is $<$99\%) and no dust extinction with $A_V<0.09$. Previously \citet{greiner11} found that a broken power-law with $A_V<0.15$ provides the best solution to the GROND-XRT SED.

%------------------TABLE RESULTS--------------------------------
\begin{table*}
%\begin{minipage}[ht]{\columnwidth}
\caption{Results of the FM best-fit parameters for the GRB afterglow X-ray to X-shooter SEDs. The columns give the burst name, the equivalent neutral hydrogen column density $N_{\rm H,X}$, optical slope $\beta_1$, X-ray slope $\beta_2$, break frequency $\nu_{\rm break}$, UV linear component intercept $c_1$, UV linear component slope $c_2$, far-UV curvature $c_4$, total-to-selective extinction $R_V$, dust content $A_V$, reduced $\chi^2$ together with number of degrees of freedom (dof) and the null hypothesis probability (NHP) for the best fit model, and F-test probability to compare single and broken-power law models. The second-last row provides the weighted mean values and 1$\sigma$ errors of all extinction curves parameters. The standard deviations around the weighted mean values are provided in the last row.}      
\label{best-fit} 
\centering
\renewcommand{\footnoterule}{}  % to avoid a line before footnotes     
\setlength{\tabcolsep}{1.5pt}
\begin{tabular}{l c c c c c c c c c c c}\hline\hline                       
GRB & $N_{\rm H,X}$ & $\beta_1$ & $\beta_2$ & log $\nu_{\rm break}$ & $c_1$ & $c_2$ & $c_4$ & $R_V$ & $A_V$ & $\chi^2_\nu$/dof (NHP\%) & F-test \\
	& 10$^{22}$ cm$^{-2}$ &  & & Hz & $\mu$m &  $\mu$m$^2$ 	& & & mag & & prob. \\ 
\hline\hline
090313	& 	4.23$^{+1.61}_{-2.10}$ & 1.18$^{+0.12}_{-0.16}$  & $\cdots$  & $\cdots$ & $-4.92\pm0.09$ & $2.10\pm0.10$ & $0.49\pm0.07$ & $2.67^{+0.13}_{-0.17}$ & $0.30\pm0.06$ & 0.92/21759(100) & $0.96$ \\
090926A	& 	0.55$^{+0.31}_{-0.18}$ & 0.82$^{+0.09}_{-0.11}$ & $\cdots$  & $\cdots$ & $\cdots$ & $\cdots$ & $\cdots$ & $\cdots$ & $<0.04$ & 0.97/28733(100) & 1.00 \\
100219A	&	$<6.67$   & $0.55^{+0.06}_{-0.07}$ & $\cdots$  & $\cdots$ & $-5.16\pm0.03$ & $2.37\pm0.05$ & $1.48\pm0.08$ & $2.65^{+0.09}_{-0.09}$ & $0.14\pm0.03$ & 1.02/17021(2.85) & 1.00 \\
100316B	& 	$<0.43$ & $0.58^{+0.08}_{-0.09}$ & $1.08^{+0.05}_{-0.04}$ & $16.67\pm0.06$ & $\cdots$ & $\cdots$ & $\cdots$ & $\cdots$ & $<0.09$ & 1.03/36438(0.04) & 0.01 \\
100418A	&	0.28$^{+0.19}_{-0.12}$ & $1.01^{+0.12}_{-0.10}$ & $\cdots$ & $\cdots$ & $-5.37\pm0.07$ & $2.30\pm0.06$ & $0.68\pm0.09$ & $2.42^{+0.08}_{-0.10}$ & $0.12\pm0.03$ & 0.86/34588(100) & 0.84 \\
100814A	& 	0.21$^{+0.10}_{-0.08}$ & $0.92^{+0.12}_{-0.08}$ & $\cdots$ & $\cdots$ & $\cdots$ & $\cdots$ & $\cdots$ & $\cdots$ & $<0.07$ & 0.98/33495(99.7) & 1.00 \\
100901A	& 	0.22$^{+0.12}_{-0.10}$ & $0.70^{+0.13}_{-0.16}$ & $1.20^{+0.11}_{-0.08}$  & $15.73\pm0.07$ & $-3.57\pm0.07$ & $2.04\pm0.05$ & $0.04\pm0.08$ & $3.01^{+0.11}_{-0.11}$ & $0.25\pm0.08$ & 1.03/31421(0.05) & $<$0.01 \\
101219B	& 	0.08$^{+0.05}_{-0.03}$  & $0.93^{+0.14}_{-0.10}$ & $\cdots$  & $\cdots$ & $\cdots$ & $\cdots$ & $\cdots$ & $\cdots$ & $<0.11$ & 0.96/41338(100) & 1.00 \\
111008A	& 	2.14$^{+1.18}_{-1.03}$ & $0.47^{+0.06}_{-0.06}$ & $0.97^{+0.05}_{-0.07}$  & $16.25\pm0.08$ & $-4.84\pm0.07$ & $2.36\pm0.05$ & $0.44\pm0.06$ & $2.47^{+0.07}_{-0.09}$ & $0.12\pm0.04$ & 0.92/22541(100) & 0.01 \\
111209A	& 	0.16$^{+0.06}_{-0.05}$ & $0.68^{+0.12}_{-0.09}$ & $1.18^{+0.08}_{-0.09}$ & $16.12\pm0.11$ & $-5.12\pm0.11$ & $2.28\pm0.08$ & $0.72\pm0.06$ & $2.53^{+0.13}_{-0.15}$ & $0.18\pm0.08$ & 1.01/38817(16.9) & $<$0.01 \\
120119A	& 	0.94$^{+0.66}_{-0.42}$ & $0.84^{+0.10}_{-0.09}$ & $\cdots$  & $\cdots$ & $-4.13\pm0.08$ & $2.09\pm0.07$ & $0.22\pm0.10$ & $2.99^{+0.24}_{-0.18}$ & $1.02\pm0.11$ & 0.91/36437(100) & 0.84 \\
120815A	&	$<0.97$ & $0.92^{+0.10}_{-0.10}$ & $\cdots$  & $\cdots$ & $-4.77\pm0.08$ & $2.14\pm0.07$ & $0.82\pm0.08$ & $2.38^{+0.09}_{-0.09}$ & $0.19\pm0.04$ & 1.02/31427(0.60) & 1.00 \\
120923A	& 	$<26.1$ & 0.91$^{+0.10}_{-0.21}$ & $\cdots$  & $\cdots$ & $\cdots$ & $\cdots$ & $\cdots$ & $\cdots$ & $<0.16$ & 0.93/5302(100) & $0.59$ \\
121024A	& 	$<1.47$  & $0.85^{+0.09}_{-0.13}$ & $\cdots$  & $\cdots$  & $-4.23\pm0.06$ & $2.20\pm0.08$ & $0.57\pm0.05$ & $2.81^{+0.20}_{-0.16}$ & $0.26\pm0.07$ & 1.02/29533(0.44)  & 0.98 \\
130427A	&	0.08$^{+0.02}_{-0.02}$ & $0.74^{+0.04}_{-0.04}$ & $\cdots$  & $\cdots$ & $-5.06\pm0.10$ & $2.24\pm0.09$ & $0.42\pm0.07$ & $2.92^{+0.19}_{-0.14}$ & $0.11\pm0.04$ & 0.94/30432(100) & 1.0 \\
%130603B  & 	0.25$^{+0.11}_{-0.08}$ & $0.70^{+0.15}_{-0.19}$ & $1.20^{+0.09}_{-0.11}$ & $16.24\pm0.11$ & $-3.78\pm0.07$ & $1.92\pm0.11$ & $0.63\pm0.05$ & $2.81\pm0.13$ & $0.85\pm0.08$ & 39753 (38963) \\
130606A  &	$<4.10$ & $0.96^{+0.05}_{-0.12}$ & $\cdots$  & $\cdots$ & $\cdots$ & $\cdots$ & $\cdots$ & $\cdots$ & $<0.07$ & 0.98/26351(99.6) & 1.0 \\
131117A  &	$<1.78$ & $0.42^{+0.11}_{-0.09}$ & $0.92^{+0.06}_{-0.08}$ & $15.95\pm0.13$ & $\cdots$ & $\cdots$ & $\cdots$ & $\cdots$ & $<0.11$ & 0.95/30278(100) & $<$0.01 \\
\hline
%$WM$ & $\cdots$ & $\cdots$ & $\cdots$ & $\cdots$ & $-4.82\pm0.19$ & $2.25\pm0.07$ & $0.58\pm0.12$ & $2.59\pm0.08$ & $\cdots$ & $\cdots$ \\
$WM$ & $\cdots$ & $\cdots$ & $\cdots$ & $\cdots$ & $-4.83\pm0.08$ & $2.23\pm0.05$ & $0.59\pm0.02$ & $2.59\pm0.07$ & $\cdots$ & $\cdots$ \\
$Stddev$ & $\cdots$ & $\cdots$ & $\cdots$ & $\cdots$ & $0.22$ & $0.15$ & $0.06$ & $0.21$ & $\cdots$ & $\cdots$ \\
\hline
\end{tabular}
%\end{minipage}
\end{table*}
%ch = [19980.,27989.,17374.,37351.,29647.,32789.,32247.,39853.,20828.,39084.,33281.,32061.,4924.,30173.,28572.,25749.,28657.]
%d = [21759.,28733.,17021.,36438.,34588.,33495.,31421.,41338.,22541.,38817.,36437.,31427.,5302.,29533.,30432.,26351.,30278.]   

\subsection{GRB\,100418A}
The X-shooter observations of GRB\,100418A ($z=0.624$) were carried out at 0.4, $\sim$1.5, and 2.4 days after the burst. The first two epochs have good flux calibration and the third epoch spectra have contamination from the GRB host galaxy and supernova emission. The NIR data of the first epoch has low S/N in the $K$-band. We, therefore, used the second epoch observations for our SED analysis due to having better NIR data. Photometric data for the SED normalisation were obtained with GROND in all seven bands from de Ugarte Postigo (in prep). We find the SED fits well with a single power-law and $A_V=0.12\pm0.03$ with a featureless extinction curve with $R_V=2.42^{+0.08}_{-0.10}$. Note that even with lesser NIR data, the dust content at the first two epochs remains $A_V\sim0.12$. \citet{deugarte11} derived intrinsic extinction of $A_V=0.09\pm0.04$ with an SMC law using the X-shooter spectra. \citet{japelj15} found the best fit with a broken power-law and $A_V=0.20^{+0.03}_{-0.02}$ with an SMC extinction curve for the X-shooter-XRT SED at a similar epoch as ours. \citep{japelj15} reported a break in the optical data. However, \citet{marshall11} find no evidence of a spectral break in the NIR-to-X-ray lightcurves after rebrightening at $\sim$0.6 days, consistent with our findings.

\subsection{GRB\,100814A}
The X-shooter spectra of GRB\,100814A ($z=1.440$) were obtained at $\sim$0.04, $\sim$0.09, and 4.1 days after the burst trigger. Due to flux calibration discrepancy at earlier two epochs, the data at 4.1 days after the burst are analysed here. Photometric data from \citet{nardini14} taken with GROND in all seven bands are used for the SED normalisation. The X-ray to the NIR SED fits well with a single power-law and no dust extinction with $A_V<0.07$. \citet{nardini14} fitted the GROND-XRT SEDs at four epochs simultaneously with a constant value of extinction and obtained $A_V<0.04$ with an SMC curve, consistent with our results. \citet{japelj15} fitted the X-shooter-XRT SED and found the best fit with a broken power-law and an SMC extinction curve with $A_V=0.20\pm0.03$. The spectral break was found within the optical data. \citet{nardini14} found a spectral break evolution by fitting SEDs at four different epochs where the latter case at $\sim$5 days is consistent with single power-law. This is consistent with our results suggesting no spectral break in the optical data.

\subsection{GRB\,100901A}
The X-shooter spectra of GRB\,100901A ($z=1.408$) were acquired at 2.758 days after the burst trigger. For SED normalisation, photometric data are taken from Gomboc et al. (in prep.). The NIR to X-ray SED fits well with a broken power-law and $A_V=0.25\pm0.08$ and prefers a flatter extinction curve with $R_V=3.01\pm0.11$. Previously, \citet{hartoog13} using the photometric SED from Gomboc et al. (in prep.) derived $A_V=0.21$. \citet{japelj15} reported a broken power-law with an SMC extinction curve provides the best fit to these data with $A_V=0.29\pm0.03$, consistent with our values.

\subsection{GRB\,101219B}
The X-shooter spectra of the afterglow of GRB\,101219BA ($z=0.552$) were taken at three epochs at 0.483, 16.4, and 39.6 days after the burst. We used the first epoch at 0.483 days after the burst for the SED analysis as the other two epoch spectra have SN contamination \citep{sparre11}. Photometric data from GROND in all seven bands are available at the first epoch \citep{sparre11} and used for the spectra normalisation. The SED provides a good fit with a single power-law and no dust extinction with $A_V<0.11$. \citet{sparre11} also found no dust extinction for the similar data with $A_V<0.1$. 

\subsection{GRB\,111008A}
The X-shooter spectra of GRB\,111008A ($z=4.991$) were taken at $\sim$0.355 days after the burst trigger. GROND photometry in the $z^\prime JH$ and K bands are available at 0.41 days after the burst from \citet{sparre14}. We normalised the X-shooter spectra to the level of the GROND photometry to fit for intrinsic extinction curve. The SED prefers a broken power-law and a featureless steep extinction curve ($R_V=2.47^{+0.07}_{-0.08}$) with $A_V=0.12\pm0.04$. \citet{sparre14} found that the GROND-XRT data fitted well with a broken power-law and an SMC extinction curve with $A_V=0.11\pm0.04$. Recently, \citet{bolmer17} suggested that the GRB photometric SED can be explained by an SMC-type curve with $A_V=0.13^{+0.03}_{-0.07}$, both findings consistent with our extinction values. 

\subsection{GRB\,111209A}
The X-shooter spectra of GRB\,111209A ($z=0.677$) afterglow were taken at 0.74 and 19.82 days after the burst. We used the first epoch spectra for the SED analysis where the Gemini-North photometric observations in the $gri$ and $z$ bands are available at 0.90 days after the burst from \citet{levan14}. After the normalisation, the SED at 0.90 days fits well with a broken power-law and a featureless extinction curve ($R_V=2.53^{+0.13}_{-0.15}$) with a visual extinction of $A_V=0.18\pm0.08$. Previously, \citet{stratta13} performed multi-epoch SED analysis and reported that the GRB afterglow can be explained with dust extinction of $A_V=0.3$--$1.5$, that may undergo dust destruction at late times.

\subsection{GRB\,120119A}\label{inbump}
The X-shooter spectra of GRB\,120119A ($z=1.728$) were obtained at $\sim$0.074 and 0.2 days after the burst trigger. We here use the X-shooter observations taken at 0.074 days after the burst due to better signal-to-noise. Photometric data for SED normalisation are obtained from \citet{morgan14} in the $BRIJH$ and $K$ bands. The SED fits well with a single power-law and featureless SMC-like extinction curve ($R_V=2.97^{+0.19}_{-0.22}$) with $A_V=1.02\pm0.11$. This burst has the highest amount of extinction in our sample. For such a high extinction usually 2175\,\AA\ bumps are seen \citep{zafar11,schady12}. Due to the redshift of the burst the location of a 2175\,\AA\ bump would be between the UVB and VIS arm data. The reduction is poor in those regions due to instrumental effect and excluding those regions, the SED fit provides a bump strength of $c_3<0.43$. Previously \citet{morgan14} found the best fit with an SMC-type extinction curve and $A_V=0.88\pm0.01$ for this burst at $\sim39$ min after the burst. Later, using the X-shooter spectrum \citet{japelj15} found that none of the standard extinction curves can fit the data well and extinction for this burst is  $A_V=1.07\pm0.03$.

\subsection{GRB\,120815A}
The X-shooter spectra of GRB\,120815A ($z=2.358$) afterglow were taken at 0.086 days after the burst trigger. The GROND afterglow photometry is available at 0.084 days after the burst in the $g^{\prime}r^{\prime}i^{\prime}z^{\prime}$ and $J$, where $g^{\prime}$ band data is affected by the Ly$\alpha$ absorption \citep{kruhler13}. We normalised our spectra to the available photometry. Our SED fits well with a single power-law and the steepest extinction curve of our sample $R_V=2.38\pm0.09$ and $A_V=0.19\pm0.04$. \citet{kruhler13} found that the GROND-X-ray SED fit prefers an SMC extinction law with $A_V=0.15\pm0.02$, consistent with our results. However, with the X-shooter spectra, \citet{japelj15} found that an SMC law with $A_V=0.32\pm0.02$ provides the best solution to these data. Such higher extinction value could be a result of using fixed law and finding a break in the optical data. In contrast, \citet{kruhler13} reported an absence of spectral break from the NIR-to-X-ray lightcurves analysis.

\subsection{GRB\,120923A}
The spectrum of the highest redshift burst (GRB\,120923A: $z=7.840$) of our sample was taken with the X-shooter at 0.82 days after the burst trigger. The X-shooter spectrum shows GRB trace redward of 1200\,nm, therefore, only NIR arm observations are used for the SED analysis. At 0.789 days after the burst, photometric data in the $JH$ and $K_s$ band were taken with the VLT Infrared Spectrometer And Array Camera (ISAAC) reported by \citet{tanvir17}. We normalised the X-shooter-XRT observations to the photometric data. The SED at 0.789 days after the burst fits well with a single power-law and no extinction with $A_V<0.16$. \citet{tanvir17} reported marginal to no dust extinction with $A_V<0.2$ from a separate photometric and spectroscopic analysis, consistent with our findings.

\subsection{GRB\,121024A}
\begin{figure}
  \centering
{\includegraphics[width=\columnwidth,clip=]{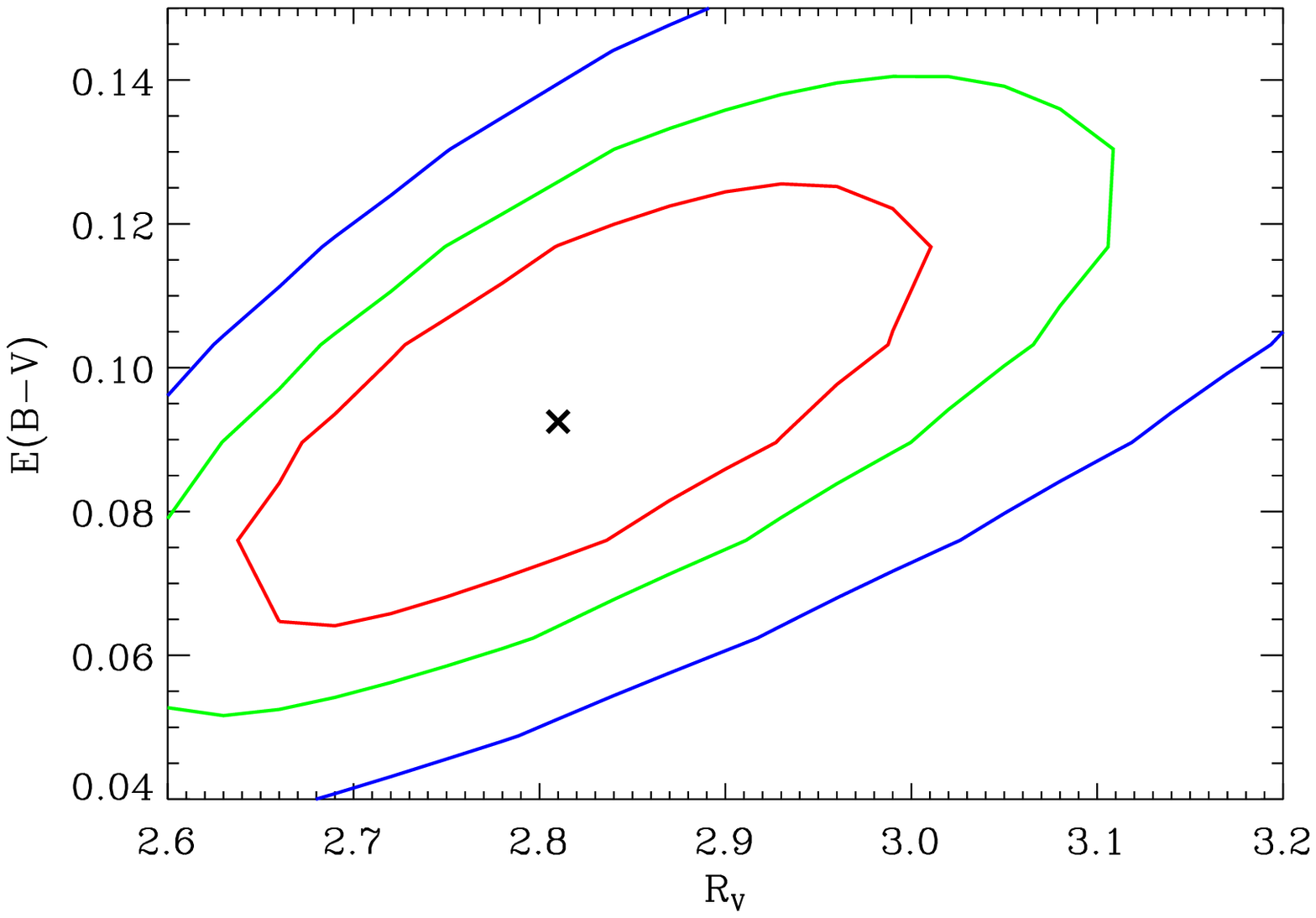}}
%{\includegraphics[width=0.495\columnwidth,clip=]{/Users/tzafar/work/GRB-X-shooter/SED_fit/121024A/xspec_fit/contourphex.ps}}
     \caption{1$\sigma$ (red), 2$\sigma$ (green), and 3$\sigma$ (blue) contours for the dust extinction, $E(B-V)$, and total-to-selective extinction, $R_V$, are shown for the case of GRB\,121024A.}
     %in the left panel and spectral index $\Gamma_1$ and $E(B-V)$ are plotted in the right panel for the case of GRB\,121024A.}
         \label{cont:121024A}
           \end{figure}

The X-shooter spectra of GRB\,121024A ($z=2.298$) were acquired at $\sim$0.075 days after the burst. Photometric data for the SED normalisation were obtained with GROND in all seven bands \citep{wiersema14,varela16}. The SED is generated at 0.128 days after the burst. The SED fits well with a single power-law and featureless extinction curve ($R_V=2.81^{+0.20}_{-0.16}$) with $A_V=0.26\pm0.07$ (see Fig \ref{cont:121024A} as an example of the quality of the fit). Previously, \citet{wiersema14} and \citet{varela16} found the best fit value of $A_V=0.22\pm0.02$ and $A_V=0.18\pm0.04$, respectively with an SMC-type extinction curve. Using the dust-to-metals correction, \citet{friis15} expected an extinction of $A_V=0.9\pm0.3$ requiring $R_V>15$. However, their GROND-XRT data fits well with an SMC extinction model with $A_V=0.09\pm0.02$.

\subsection{GRB\,130427A}
The X-shooter spectra of GRB\,130427A ($z=0.339$) were taken at 0.52 days after the burst \citep{xu13}. The photometric data in the $g^{\prime}r^{\prime}i^{\prime}z^{\prime}JH$ and $K$ bands at $\sim$0.65 days after the burst are taken from the lightcurves provided by \citet{perley14}. The SED is scaled to the photometry level at 0.65 days. The SED provides a good fit with a single power-law and a featureless extinction curve ($R_V=2.92^{+0.19}_{-0.14}$) with a small amount of extinction $A_V=0.11\pm0.04$. \citet{perley14} performed radio to the 100 GeV Fermi Large Area Telescope (LAT) SED fits at different epochs and reported that the optical/NIR data prefers an SMC extinction curve with $A_V=0.13\pm0.06$, consistent with our results. However, \citet{japelj15} suggested an SMC curve with $A_V=0.16\pm0.02$ and a broken power-law (with $\Delta\beta=0.31\pm0.05$) with the break in the NIR data could explain the SED at 0.7 days. By leaving slope difference $\Delta\beta$ as a free parameter, we still find that the SED is consistent with a single power-law and there is no evidence of a shallower optical slope. The lower S/N of the NIR spectrum and further coverage hinder us to see such an optical break. Note that the $J$ and $H$ band data are consistent with a single power-law. 

%Moreover, normalisation between the different arms could mimic presence of an optical break.
%\subsection{GRB\,130603B}
%This is a short-GRB in our sample. The X-shooter spectra of this GRB are taken at 0.356 days after the burst trigger. The optical photometry at 0.354 days after the burst is taken from \citet{deugarte14}. The X-shooter SED normalised to the level of the photometry provides best results with an SMC-like extinction curve and $A_V=0.85\pm0.08$. Previously, \citet{deugarte14} found the similar data can be explained by an SMC extinction curve and $A_V=0.86\pm0.15$. \citep{japelj15} reported the SED prefers and SMC extinction curve with an $A_V=1.19^{+0.23}_{-0.12}$, all suggesting similar results.

\subsection{GRB\,130606A}
The X-shooter spectra of GRB\,130606A ($z=5.913$) afterglow were taken at 0.33 days after the burst. At $\sim$0.329 days after the burst trigger, the GROND photometric data in the $z^\prime JH$ and $K$ band from GROND are taken from \citet{afonso13}. The X-shooter data is scaled to the photometry. The NIR to the X-ray SED fits well with a single power-law and no dust extinction with $A_V<0.07$. Previously \citet{hartoog15}, \citet{japelj15}, and \citet{bolmer17} also found no dust extinction for this burst with  $A_V<0.2$.

\subsection{GRB\,131117A}
The X-shooter spectra of GRB\,131117A ($z=4.042$) afterglow were taken at $\sim$0.05 days after the burst trigger. At $\sim$0.02 days after the burst, the GROND photometric data in the $i^\prime z^\prime J$ and $H$ band from GROND were taken from \citet{bolmer17}. The X-shooter data is scaled to the photometry at $\sim0.02$ days. The NIR to the X-ray SED provides a good fit with a broken power-law and no dust extinction with $A_V<0.11$. \citet{bolmer17}reported no dust extinction for this burst with $A_V<0.09$, consistent with our findings.

\subsection{X-ray analysis}
In Table \ref{best-fit}, the equivalent hydrogen column densities ($N_{\rm H,X}$) derived through the simultaneous X-shooter to X-ray SED analysis for each GRB case is provided. For our time sliced spectra, we obtained significant $N_{\rm H,X}$ measurements for 10 cases only. We compared the $N_{\rm H,X}$ values with the redshift of GRBs. For our smaller sample, we find an evolution of the $N_{\rm H,X}$ with redshift as suggested by \citet{campana12}. For measurements only, the statistical analysis results in a Spearman rank correlation coefficient $r=0.89$ and $>99$\% probability. The slope of the correlation is $0.40\pm0.04$. This is in contrast with the findings of \citet{buchner17} suggesting no evolution of the $N_{\rm H,X}$ with redshift. Comparing $N_{\rm H,X}$ values of our sample with \citet{buchner17} indicates their values are up to $1.0$\,dex smaller.

%----------------------------------------------------------------
\section{Discussion}
\subsection{Average GRB extinction curve}\label{avg:ext}
\begin{figure}
  \centering
{\includegraphics[width=\columnwidth,clip=]{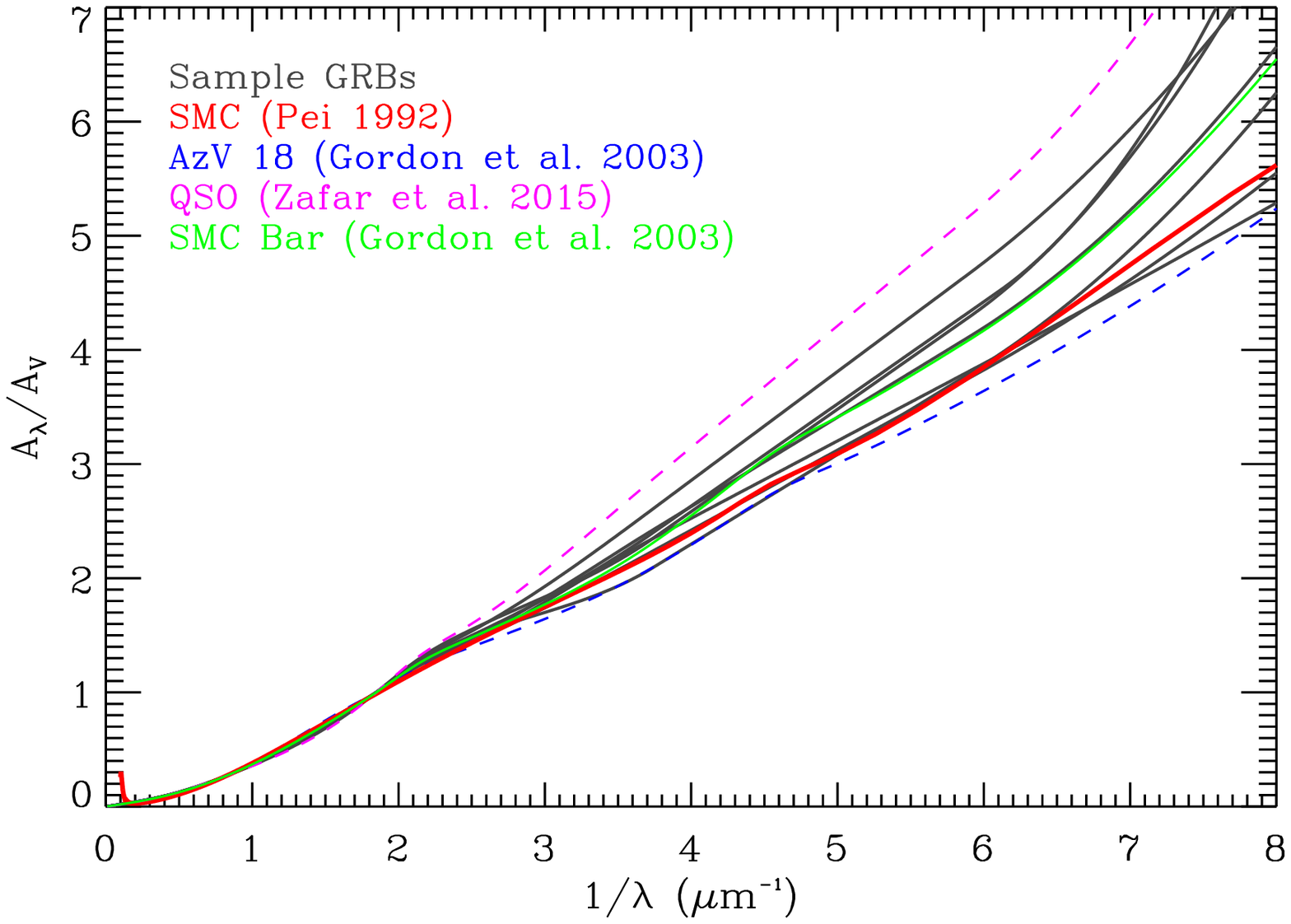}}
     \caption{Individual extinction curves of 7 GRB afterglows with broadest wavelength coverage in our sample are compared with the extinction curve of the average SMC (red) taken from \citet{pei92}. For a comparison, the blue, green, and magenta dashed curves show the flattest SMC extinction curve towards sightline AzV\,18, SMC Bar extinction curve \citep{gordon03}, and steepest QSO extinction curve \citep{zafar15}, respectively. GRB extinction curves derived in this work usually appear to be steeper than the typical SMC curve, however, have a broad range of $R_V$ values. }
         \label{grb:sample}
           \end{figure}

Seven GRBs in our sample are consistent with no dust extinction and their 3$\sigma$ $A_V$ upper limits are provided in Table \ref{best-fit}. 10 extinguished out of 17 GRBs have small to high extinction values with $A_V$ values ranging from $0.11\pm0.04$ to $1.02\pm0.11$ resulting in a mean of $\langle A_V\rangle = 0.27$\,mag (standard deviation of 0.27). \citet{covino13} found from a sample of 58 GRBs that 50\% prefer less than 0.3--0.4\,mag extinction. In our smaller sample, 94\% bursts have $\lesssim0.3$\,mag extinction. Individual best-fit extinction curves are shown as black full lines in Fig. \ref{grb:sample}, but as explained below we do not here include the three lowest redshift objects due to the shortened coverage in rest-frame UV.  For comparison we also plot several standard extinction curves:  the SMC extinction
curve from \citet{pei92}, the flattest known SMC extinction curve (with $c_1=-4.94\pm0.63$, $c_2=2.27\pm0.20$, $c_4=0.18\pm0.08$, and $R_V=3.30\pm0.38$) towards sightline AzV\,18 \citep{gordon03}, the SMC Bar extinction curve ($c_1=-4.96\pm0.20$, $c_2=2.26\pm0.04$, $c_4=0.46\pm0.08$, and $R_V=2.74\pm0.13$) from \citet{gordon03}, and the steepest intrinsic QSO extinction curve \citep{zafar15} derived for a sub-sample of High $A_V$ quasar (HAQ; \citealt{krogager15}) survey. Out of 10 dusty cases in our sample, three GRBs (GRB\,100418A, GRB\,111209A, and GRB\,130427A) were at low redshifts ($z$$<$0.7) and are therefore lacking rest-frame UV information of their extinction curves. Of the remaining seven cases, three GRB afterglows prefer extinction curves steeper than the
typical \citet{pei92} SMC curve (the three upper black curves in Fig. \ref{grb:sample}). However, $R_V$ values rather than clustering around a particular value show a remarkably even distribution of $R_V$ values bracketed by the AzV\,18 and the HAQ curves. In particular, the histogram of $R_V$ values of seven GRBs with broad coverage has a flat distribution between values ranging from 2.38 to 3.01 (see Table \ref{best-fit}).

\begin{figure}
  \centering
{\includegraphics[width=\columnwidth,clip=]{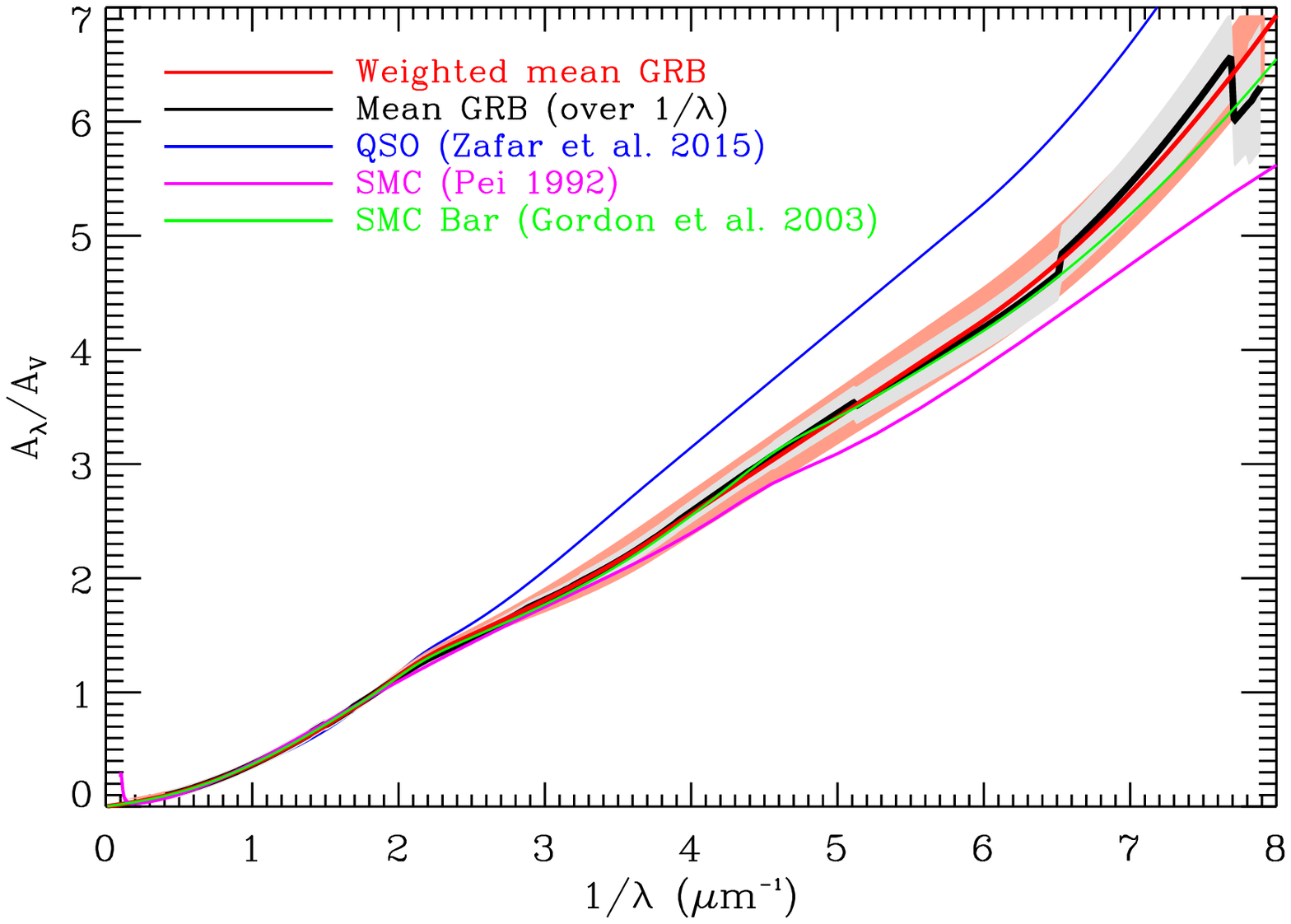}}
     \caption{`Average GRB extinction curves' computed using the $WM$ values of seven extinguished GRBs (red) and by estimating the mean for each 1/$\lambda$ ($\mu$m$^{-1}$) over the X-shooter observational coverage (black). The red and grey shaded areas represent the combined 1$\sigma$ error on the averages (see \S\ref{avg:ext} for more details on the computation methods). The magenta and green solid lines indicate the typical SMC \citep{pei92} and average SMC Bar \citep{gordon03} extinction curves, respectively. The average steep QSO extinction curve from \citet{zafar15} is shown in blue solid line for a comparison.}
         \label{mean:grb}
           \end{figure}

To derive the {\it `average GRB extinction curve'} for our sample, we adopt
two methods. First method is to generate the {\it `average GRB extinction curve'} by calculating the {\it weighted mean} ($WM$) values of the best-fit parameters.
It is seen from the $R_V$ measurements in Table \ref{best-fit} that the scatter of the distribution is much larger than the individual measurement errors, i.e. the scatter represents a real intrinsic distribution. For this reason we provide,  in the bottom two
rows of Table \ref{best-fit}, both the $WM$ and its corresponding error, as well as the measured scatter ($Stddev$) of the distribution. As before, we again discard the three low redshift GRBs (GRB\,100418A, GRB\,111209A, and GRB\,130427A
and re-compute the WM values and corresponding errors. This results in $c_1=-4.76\pm0.09$, $c_2=2.21\pm0.05$, $c_4=0.58\pm0.02$, and $R_V=2.61\pm0.08$, identical to the values listed in Table \ref{best-fit}) to within less than 1$\sigma$ errors for all parameters. I.e., the final curve is not strongly influenced by those three objects. We then use these $WM$ values to draw the red colour extinction curve in Fig. \ref{mean:grb}, where the red shaded area marks the combined uncertainty via propagation of errors on all parameters. 

In our second method, we compute the {\it `average GRB extinction curve'}. i.e.
the average of the actual individual curves rather than the parameters, but
here only taking into account the 1/$\lambda$ ($\mu$m$^{-1}$) wavelength range 
covered by each X-shooter observation. For each rest-wavelength 1/$\lambda$ ($\mu$m$^{-1}$) a 
mean and 1$\sigma$ error on $A_\lambda/A_V$ is computed. Because of the vastly
different redshifts, the resultant mean extinction curve must display a series
of `steps' where each spectrum begins and ends. That curve 
is shown in Fig. \ref{mean:grb} in black colour. The grey shaded area illustrates
the 1$\sigma$ error on the extinction curve. As can be seen from the figure the
agreement between the results of the two methods is excellent.  Hereafter, we
shall refer to the $WM$ extinction curve as the
{\it `average GRB extinction curve'}. From Fig. \ref{mean:grb} it is also seen that the {\it average GRB extinction curve} is slightly steeper than the canonical
\citet{pei92} SMC curve (with $R_V=2.93$) but matches well the SMC Bar extinction
curve presented in \citet{gordon03}.

%The {\it average GRB extinction curve} is slightly steeper than the canonical \citet{pei92} SMC curve (with $R_V=2.93$) but matches the SMC sightline AzV\,23 ($c_1=-5.17\pm0.29$, $c_2=2.38\pm0.09$, $c_4=0.46\pm0.06$, and $R_V=2.65\pm0.17$) or the SMC-bar extinction curve ($c_1=-4.96\pm0.20$, $c_2=2.26\pm0.04$, $c_4=0.46\pm0.08$, and $R_V=2.74\pm0.13$) presented in \citet{gordon03}. }

Note that our sample contains five GRBs above $z>4$. Only two of them have significant but small amount of dust (GRB\,100219A: $A_V=0.14\pm0.03$ and GRB\,111008A: $A_V=0.12\pm0.04$) while three (GRB\,120923A: $A_V<0.16$, GRB\,130606A: $A_V<0.07$, and GRB\,131117A: $A_V<0.11$) are consistent with no dust. These high redshift cases are consistent with the findings of a decrease in dust content at $z>4$ suggested by \citet{zafar11b} and recently confirmed by \citet{bolmer17}.
 
\subsection{Extinction curves comparison}
None of our extinguished GRBs have a significant 2175\,\AA\ dust feature (but see
Sect. \ref{inbump}).
In Fig. \ref{mean:grb} we plot a few featureless comparison extinction
curves from the SMC and from QSOs. We use Kolmogorov-Smirnov statistics to
determine the significance of the difference between the various extinction curves
and find that the average intrinsic QSO extinction
curve, derived for a HAQ survey sub-sample, from \citet{zafar15} with
$R_V=2.2\pm0.2$ deviates from the average GRB extinction curve
at $>$99\% confidence level. The average GRB curve is similar to but slightly steeper than the typical \citet{pei92} SMC extinction curve. It is consistent with the SMC Bar extinction curve from \citet{gordon03} at the $\sim95$\% confidence level. We find that the SMC Bar extinction curve (the green curve in Fig. \ref{grb:sample} and \ref{mean:grb}) is on average a better representation of our sample.

\citet{wiersema11} fitted hydrogen and helium fluxes of several recombination transitions seen in the spectra of GRB host galaxies to estimate $R_V$ using the \citet{cardelli89} parametrisation. The extinction over all recombination line regions combined, rather than one sightline (as in this work) resulted in $R_V$ higher than the average MW value. Previously, \citet{schady12} attempted to fit the parametric extinction curve to the GRB X-ray to optical/NIR photometric data. For moderately extinguished GRBs ($A_V<1$), their best fit \citet{fm90} extinction curve is closer to the LMC ($R_V=3.16$). Although due to the absence of a 2175\,\AA\ bump, the extinction model around the bump location is closer to the featureless SMC curve. 12\% of the 49 GRBs in their sample are significantly dusty with $A_V>1$ and found to have extinction curves flatter than the mean MW extinction curve. In our sample, we only have GRB\,120119A with $A_V\gtrsim1$, where the bump location falls between the edges of the UVB and VIS arms. Therefore, we are not able to determine whether or not a bump is present. For the moderate extinction cases, we also find that the GRB SEDs are better described by featureless extinction curves.

A 2175\,\AA\ bump extinction curve with steep UV slope was found for GRB\,080605 \citep{zafar12}. For GRB\,140506A, \citet{fynbo14} reported an extremely steep UV extinction curve that could be fitted by a \citet{fm07} parametrisation of an extreme 2175\,\AA\ bump, albeit with fitting parameters unlike any derived for MW sight-lines. \citet{heintz17} later excluded this possibility and the extinction curve of GRB\,140506A must hence be very steep for other reasons. \citet{zhou06} reported a steeper extinction law for a sample of $\sim$2000 UV-deficient narrow line Seyfert galaxies. \citet{leighly14} found a steep extinction law with $R_V=2.74$ for the Seyfert galaxy Mrk\,231 using the equation from \citet{goobar08}. \citet{gallerani10} found for 3.9 $<$ $z$ $<$ 6.4 QSOs that their extinction curves deviate from the SMC law and flatten at $\lambda_{\rm rest}<2000$\,\AA . In contrast, for a sample of QSOs at $z\sim6$, \citet{hjorth13} suggested that the median extinction curve is consistent with the featureless SMC curve. Recently for the HAQ survey sub-sample at 0.7 $<$ $z$ $<$ 2.1, \citet{zafar15} fitted a parametric extinction law and found that an SMC law is inadequate to define SEDs of those dusty QSOs. Their entire sample is well fit with a single best-fit value of $R_V = 2.2\pm0.2$. An even steeper extinction curve with $R_V=1.4$ is proposed for the Type Ia supernova SN\,2014JA in the starburst galaxy M82 \citep{amanullah14}, remarkably similar to that derived for GRB\,140506A \citep{heintz17}. \citet{zelaya17} found that weighted average $R_V$ of ten Type Ia SNe is $R_V=1.5\pm0.06$. They further find higher continuum foreground polarisation (by dust scattering) and hence $A_V$ for low $R_V$ values. Circular polarisation intrinsic to the source or by dust scattering effects along the line of sight is found for GRB\,121024 \citep{wiersema14}. We here have a direct measure of the $R_V$ values in young star-forming galaxies, finding slightly steeper extinction curves but no evidence for very low $R_V$.

%\citet{poznanski12} reported $E(B-V)$ and \ion{Na}{i} equivalent width correlation for SNe suggesting denser gas- and dust-rich star forming environments creating larger grains leading to large values of $R_V$.

Exploiting the broad wavelength coverage of X-shooter we are here able to fit the individual extinction curves of GRBs for the first time. The GRB afterglows have a continuum of extinction curves and are on average slightly steeper than the canonical SMC.

\subsection{Dust composition}
Our GRB extinction curves do not have a significant 2175\,\AA\ dust feature implying an absence of carbonaceous grains. Using the X-shooter data, we have enough coverage available to detect a significant 2175\,\AA\ bump but we did not discover any. It is possible that these GRB environments have small carbonaceous grains but that the steepness of the curve dilutes the feature. The steep extinction curve indicates the presence of small grains which could be due to the destruction of dust by the harsh environment \citep{reach00} of the GRB or due to GRB stellar environments being young age and/or of lower metallicities. For the first scenario, dust destruction in the ISM could be caused by sputtering, evaporative heating, and/or grain charging where destruction rate depends on grain size and composition. \citet{fruchter01} modelled the influence of different effects on dust in GRB environments and find that grain shattering could reduce the total extinction for $\lambda<1400$\,\AA . \citet{perna03} simulated dust destruction due to thermal sublimation of the dust grains resulting in a significant reduction of the 2175\,\AA\ bump and as the destruction proceeds, the extinction curve rather becomes flatter. For the second scenario of small grain presence, we looked in our sample where $R_V$ values and metallicities are reported. We find that metallicities are available only for five dusty GRBs in our sample: GRB\,100219A ($[M/H]=-1.1\pm0.2$; \citealt{thoene13}), GRB\,111008A ($[M/H]=-1.7\pm0.1$; \citealt{sparre14}), GRB\,120119A ($[M/H]=-0.79\pm0.25$; \citealt{wiseman17}), GRB\,120815A ($[M/H]=-1.15\pm0.12$; \citealt{kruhler13}), and GRB\,121024A ($[M/H]=-0.7\pm0.1$; \citealt{friis15}). Although low number statistics, comparing metallicities with $R_V$ indicates that data are linearly correlated with a coefficient $r=0.73$. A larger sample of metallicities and total-to-selective extinction values will better infer the relation between both quantities.

\citet{mathis96} proposed that steeper extinction curves could arise due to the presence of silicate grains. $\sim$70\%\ of the interstellar dust core mass is composed of silicate grains \citep{draine03}. For the cases where both a 2175\,\AA\ bump and UV steep rise is present, \citet{weingartner01} and \citet{li01} developed a model to explain the 2175\,\AA\ bump through carbonaceous grains and silicate population responsible for the steep rise. Recently, \citet{mishra17} modelled that far-UV extinction could be due to a population of both small carbon and silicates grain. The steep extinction curves observed in this work could be due to such small grains.
   
%________________________________________________________________
%Conclusions
\section{Conclusions}
We present here a sample of X-shooter selected GRB afterglow SEDs from March 2009 to March 2014. We analyse the NIR to X-ray SEDs of 17 GRB afterglows with redshifts ranging from 0.34 $<$ $z$ $<$ 7.84. We have only included GRBs for which nearly simultaneous photometry is available allowing us to proper flux calibrate the X-shooter data. Single and broken (with a change in slope fixed to $\Delta\beta=0.5$) power-laws are used to fit the SEDs together with the analytical FM extinction law. 10 out of the 17 GRBs are found to be extinguished with $A_V$ values ranging from $A_V=0.1-1.0$\,mag with a mean of $\langle A_V\rangle=0.27$\,mag. Our individual $R_V$ values have a flat distribution with values varying from 2.38--3.01. The flat distribution on $R_V$'s indicates that GRBs have a wide range of grain sizes and compositions, but on average favouring a somewhat steeper reddening than the standard SMC. We derived the `average GRB extinction curve' by obtaining the weighted mean values of the best-fit parameters of seven individual curves (chosen to have the broadest wavelength coverage) resulting in an optimal weighted combined value of $R_V=2.61\pm0.08$. This extinction curve is similar to, but slightly steeper than, the typical \citet{pei92} SMC curve.  It is fully consistent with the SMC Bar from \citet{gordon03} at $\sim95$\% confidence level. Steeper extinction curves have been previously reported for GRBs, Seyfert galaxies, Supernovae, and QSOs. Steep extinction curves are thought to be representative of a population of silicates producing small size dust grains. Large dust grains may be destroyed in harsh environments of GRBs, resulting in steep extinction curves. Another possibility could be young age and/or lower metallicities of GRBs, although statistics are low. 

%________________________________________________________________
%Acknowledgments
\section*{Acknowledgements}
The X-ray data for this work is obtained from the UK \emph{Swift} Science Data Center at the University of Leicester. KEH acknowledges support by a Project Grant (162948-051) from The Icelandic Research Fund. AdUP acknowledges support from the Spanish research project AYA 2014-58381-P, from a RyC fellowship and from a BBVA Foundation grant for Researchers and Cultural Creators. JJ acknowledges support from NOVA and NWO-FAPESP grant for advanced instrumentation in astronomy.

%==================================================

\bibliographystyle{aa}
\bibliography{grb-xsh.bib}{}

\bsp

\label{lastpage}
\end{document}